\documentclass[aps,pra,twocolumn,groupedaddress, amsmath,amssymb]{revtex4-2}

\usepackage{graphicx}
\usepackage[linkcolor=blue]{hyperref}
\usepackage{dcolumn}
\usepackage{bm}
\usepackage{xcolor}
\begin{document}


\title{Coherence from multiorbital tunneling ionization of molecules}


\author{C. H.  Yuen}
\email[]{iyuen@phys.ksu.edu}
\author{C. D.  Lin}
\email[]{cdlin@phys.ksu.edu}
\affiliation{J. R. Macdonald Laboratory, Department of Physics, Kansas State University, Manhattan, Kansas 66506, USA}

\date{\today}

\begin{abstract}
We present a simple and general coherence model for multiorbital tunnel ionization of molecules, which we incorporate into our previously developed density matrix approach for sequential double ionization [Yuen and Lin, Phys. Rev. A 106, 023120 (2022)]. 
The influence of this coherence is investigated through simulations of single ionization and sequential double ionization of N$_2$ and O$_2$ using few-cycle near-infrared laser pulses.
In the case of single ionization,  our results reveal the crucial role played by this coherence in generating population inversion in N$_2^+$, suggesting a potential mechanism for air lasing. 
Regarding sequential double ionization, we observe only minor changes in the kinetic energy release spectra when the coherence is included,  while noticeable differences in the angle-dependent dication yield for both N$_2$ and O$_2$ are found.
Based on these findings, we recommend the inclusion of multiorbital tunnel ionization coherence in models for single ionization of general molecules,  while suggesting that it can be safely neglected in the case of sequential double ionization.
\end{abstract}

\maketitle

\section{Introduction}
Multiorbital tunnel ionization (TI) of molecules is ubiquitous in strong field and ultrafast science.
It has been experimentally demonstrated that an electron can be tunnel ionized from multiple molecular orbitals to form a superposition of ionic states.
This phenomenon has been observed not only in double ionization~\cite{voss2004, wu2011, wu2012} but also in high-harmonic generation~\cite{mcfarland2008, smirnova2009, farrell2011, He2022}, strong-field dissociation~\cite{akagi2009, de2011},  and air lasing~\cite{liu2013, xu2015, yao2016, kleine2022}.

Existing theoretical approaches for multiorbital TI typically focus on ionization yields from different orbitals to obtain the total ionization yield.
However,  populations of the ionic states after the laser pulse are very often unknown since the nascent ionic states could be further coupled by the laser.
Such post-ionization dynamics could be modeled using first principle approaches,  but they are expensive computationally as the simulation box required should be large enough to contain the ionized electron in order to project the ionic channels out properly.
Alternatively,  some studies solved the time-dependent Schr\"{o}dinger equation of the ion to describe the post-ionization dynamics~\cite{smirnova2009,xu2015,yao2016,He2022}.
However,  as one neglects the ionized electron in the dynamics,  the residue molecular ion becomes an open system and different ionic states are not fully coherent.
Therefore, a density matrix approach is more appropriate for describing the interaction between the residue ion and the laser field.

Recently,  density matrix approaches for single ionization~\cite{zhang2020} and for sequential double ionization (SDI)~\cite{yuen2022,  yuen2023} of molecules have been developed,  allowing for the simultaneous consideration of TI and laser couplings.
However, these approaches neglect the coherence arising from multiorbital TI,  with the evolution of coherence being solely driven by laser couplings. 
As a result, the complete description of coherence between ionic states and between dication states are lacking.
While the simulated observables for SDI of N$_2$~\cite{yuen2022} and O$_2$~\cite{yuen2023} agree well with the experiments~\cite{voss2004,wu2011},  
the role of TI coherence in the intense laser-molecule interaction remains unclear.

In contrast,  coherence from TI of noble gas atoms has been studied both theoretically and experimentally.
Rohringer and Santra~\cite{rohringer2009} developed a time-dependent configuration-interaction-singles (TDCIS) method for calculating population and coherence of different spin-orbit states from TI of heavy noble gas atoms.
Their theoretical results demonstrated good agreement with the seminal experiment by Goulielmakis \textit{et al.}~\cite{goulielmakis2010} conducted on krypton.
On the other hand,  a simple coherence model based on the strong field approximation was proposed by Pabst \textit{et al.}~\cite{pabst2016} for TI of atoms, 
which showed good agreement with the results obtained using the TDCIS method for artificial atomic systems. 
The approach by Pabst \textit{et al.} was subsequently applied to O$_2$ by Xue \textit{et al.}~\cite{xue2021,  xue2022}.
However,  it remains uncertain whether a coherence model developed for atoms can be directly extended to molecules, 
particularly considering different orientations of a molecule with respect to the laser polarization.

The coherence between ionic states in molecules is not only a topic of fundamental interest but also drives the charge migration phenomena. 
After the removal of an electron, the residual ion can exist as a superposition of states, and the coherence between these states leads to electron density migration across the molecular skeleton~\cite{cederbaum1999}.
By monitoring coherence, it is possible to observe real-time electronic motion in a molecule. 
Moreover,  decoherence between ionic states can result in permanent charge transfer within a molecule,  offering significant potential for controlling chemical reactivity~\cite{lepine2014}.

Our recent work demonstrated that SDI can probe changes in coherence between pumped ionic states due to the nuclear dynamics of the ion~\cite{yuen2023b}.
When an intense few-cycle IR pulse arrives at some time-delay,  the pumped and nascent ionic states are coupled by the laser field. 
As the coherence between pumped ionic states changes by the nuclear motion,  the strength of the laser coupling between the ionic states changes accordingly,  leading to different populations of the intermediate ionic states. 
The intermediate ionic states are then further tunnel ionized by the probe pulse.
Consequently,  the yields of the dication are influenced by the population of the intermediate ionic states,  affecting observables such as kinetic energy release (KER) and branching ratios of dissociative dications. 
Therefore,  coherence between the pumped ionic states will be imprinted in the observables, 
and a comprehensive characterization of coherence during intense laser-molecule interaction is crucial for utilizing the SDI process as a probe.

In this article,  we propose a simple and general model for the build up of coherence from TI of molecules.
We implement the coherence model into the density matrix approach for SDI (DM-SDI)~\cite{yuen2022,  yuen2023} and apply it to investigate single and double ionization of N$_2$ and O$_2$ by a few-cycle intense IR pulse.
In the case of single ionization,  we found that TI coherence enhances the laser couplings between ionic states,  resulting in  approximately 40\% changes in the ionic population.
Particularly,  we provide evidence for the population inversion of the $B^2\Sigma_u^+$ state over the $X^2\Sigma_g^+$ state of N$_2^+$ for a wide range of alignment angles.
For SDI,  TI coherence leads to only about 20\% changes in the KER spectra of $\rm N^+ + N^+$ and $\rm O^+ + O^+$,  well within typical experimental uncertainties.
We predict that angle-dependent dication yields would exhibit signatures of TI coherence,  which can be tested against pump-probe experiments using rotational wave packets.

This article is organized as follows: In the next section,  we discuss the modeling of coherence arising from TI and provide an overview of the DM-SDI model for N$_2$ and O$_2$.
In Sec. ~\ref{sec3},  we examine the role of TI coherence in single ionization of N$_2$ and O$_2$ and suggest potential experiments to verify our results.
The influence of TI coherence on SDI of N$_2$ and O$_2$ and possible experimental observations are discussed in Sec. ~\ref{sec4}.
Finally,  we present a summary and outlook in Sec. ~\ref{sec5}.

\section{Theoretical approach~\label{sec2}}
\subsection{Coherence model}
In this section,  we discuss and develop the model for coherence build up from multiorbital TI.
First,  let's consider the electronic wave function of the ion and the outgoing electron as
\begin{align*}
| \Psi(t) \rangle = \sum_{i,k} c_{ik}(t) |i\rangle |k \rangle,
\end{align*}
where $|i\rangle$ is the electronic state of the ion,  and $|k \rangle$ is the time-independent continuum wave function for the outgoing electron,  which is discretized for convenience in notation.
The coefficient $c_{ik}(t)$ captures the change in photoelectron momentum due to the laser field.
Then,  the full density matrix $\hat{\rho}$ for the ion--electron system can be written as
\begin{align*}
\hat{\rho}(t) = \sum_{i,k} \sum_{j,k'} c_{ik}(t) c_{jk'}^\ast (t) |i\rangle |k \rangle   \langle j| \langle k' |,
\end{align*}
where one identifies $\rho_{ik,  jk'}= c_{ik}(t) c_{jk'}^\ast (t)$.

Since sequential double ionization is driven solely by the laser field,  one can simplify the theory by neglecting the dynamics of the ionized electron.
To achieve this,  one traces out the full density matrix $\rho$ over $k$,  obtaining the reduced density matrix (RDM) for the ion,
\begin{align*}
\rho_{ij}(t) = \sum_k c_{ik}(t) c_{jk}^\ast (t).
\end{align*}
Note that when $i=j$,  $\rho_{ii} = \sum_k |c_{ik}|^2$ represents the population of the $i$th ionic state.
Expressing $c_{ik} = |c_{ik}| e^{i \phi_{ik}}$,  one finds the relation 
\begin{align*}
\left| \rho_{ij}(t) \right| &= \left| \sum_k |c_{ik}(t)| |c_{jk} (t)| e^{i (\phi_{ik}-\phi_{jk})} \right| \\
& \leq \left|\sum_k |c_{ik}(t)| |c_{jk} (t)|  \right|\\
&\leq \sqrt{ \sum_k |c_{ik}(t)|^2 \sum_k |c_{jk} (t)| ^2} \\
&= \sqrt{\rho_{ii}(t) \rho_{jj}(t)},
\end{align*}
where the Cauchy-Schwarz inequality is applied.
Therefore,  expressing the total ionization rate to the $i$th state $\Gamma_i$ as the sum of partial ionization rate $\Gamma_{ik}$,  $\Gamma_i = \sum_k \Gamma_{ik}$,  modeling the coherence in a form similar to Refs.~\cite{pabst2016, xue2021}, 
\begin{align}
\rho_{ij}(t) = \int_{-\infty}^t \sqrt{\Gamma_i (t') \Gamma_j(t')} e^{-i (E_i - E_j)t'} e^{i \phi_{ij}(t')} dt'
\label{eq:others_model}
\end{align}
for some phases $\phi_{ij}$ may overestimate the coherence in the case of molecules due to the above inequality.

Although more accurate coherence models based on strong field approximation,  semi-classical approaches~\cite{arnold2020},  
or first principle approaches~\cite{rohringer2009} could be extended to multiorbital TI of molecules, 
they can be computationally expensive when considering different molecular orientations.
Therefore,  it is preferable to have a simple and general coherence model for multiorbital TI of molecules.

An alternative way to describe ion--electron wave function is through the partial wave expansion,
\begin{align*}
| \Psi(t) \rangle = \int \sum_{i,\nu} c_{i,E \nu}(t) |i\rangle |E \nu \rangle \,  dE,
\end{align*}
where $\nu$ is the collective indexes of partial waves in the parabolic coordinates~\cite{tolstikhin2011} and $E$ is the energy of the ionized electron.
Then,  the RDM of the ion can be expressed as
\begin{align*}
\rho_{ij}(t) = \int \sum_{\nu} c_{i, E \nu}(t) c_{j,  E \nu}^\ast (t) \,  dE.
\end{align*}

To simplify the modeling,  several assumptions are made.
Firstly,  similar to the strong field approximation,  it is assumed that the momentum of the ionized electron is $\vec{k}_i(t) = \vec{p}_i + \vec{A}(t)$ after TI,  
where $\vec{p}_i$ is the momentum after the tunnel exit,  and $\vec{A}(t)$ is the vector potential of the laser field.
Consequently,  the RDM for the ion can be obtained by tracing over $\vec{p}$ instead of $\vec{k}$. 
Additionally, it is assumed that the kinetic energy distribution $f(E)$ of the ionized electrons from different orbitals at the tunnel exit are identical.
This assumption leads to the approximation,
\begin{align*}
c_{i, E \nu}(t) \approx c_{i, \nu}(t) \sqrt{f(E)},
\end{align*}
with $\int f(E) dE = 1$.
Then,  the RDM of the ion can be approximated as
\begin{align*}
\rho_{ij}(t) \approx \sum_{ \nu} c_{i, \nu}(t) c_{j,  \nu}^\ast (t).
\end{align*}

At the lowest order of the weak field asymptotic theory,  $\nu \approx m$~\cite{tolstikhin2011},  which corresponds to the magnetic quantum number.
Therefore,  at the lowest order of the tunneling theory,  the RDM of the ion is approximated as
\begin{align*}
\rho_{ij}(t) \approx \sum_{m} c_{i,  m}(t) c_{j,  m}^\ast (t),
\end{align*}
suggesting that the evolution of coherence can be modeled as
\begin{align}
\rho_{ij}(t) = \int_{-\infty}^t \sum_m \gamma_{im} (t') \gamma^\ast_{jm}(t') e^{-i (E_i - E_j)t'} dt',
\label{eq:motif}
\end{align}
where $\gamma_{im}$ is partial ionization amplitude with $|\gamma_{im}|^2 = \Gamma_{im}$,  which is the partial ionization rate.

Since the lowest order of the weak field asymptotic theory is similar to the molecular Ammosov-Delone-Krainov (MO-ADK) theory,  
using the adiabatic approximation,
one identifies $\gamma_{im}$ as~\cite{tong2002, tolstikhin2011}
\begin{align}
\gamma_{im}(t) &= \frac{B_{im}(t)}{\sqrt{2^{|m|}|m|!}} \frac{1}{\kappa^{Z/\kappa_i - 1/2}}  \left(\frac{2\kappa_i^3}{|F(t)|}\right)^{Z/\kappa_i - (|m|+1)/2} \nonumber \\
& \times  \exp{\left[ \frac{-\kappa_i^3}{3|F(t)|} + \frac{i\pi}{4} + i\pi \left( \frac{Z}{\kappa_i} - \frac{|m|+1}{2} \right)\right]},
\label{eq:gamma}
\end{align}
where $\kappa = \sqrt{2I_p}$ with $I_p$ being the ionization potential and $Z$ being the effective charge after ionization.
For field $F(t)>0$ and $F(t)<0$,  $B_{im}(t)$  takes the form,
\begin{align}
B_{im}(t) =
\begin{cases}
 \sum_{lm'} C_{i, lm'} \left( D^l_{m' m}(\hat{R})\right)^\ast Q(l,m),  \\
 \sum_{lm'} (-1)^{l-m'} C_{i,  lm'} \left( D^l_{m' m}(\hat{R})\right)^\ast Q(l,m), 
\end{cases}
\label{eq:struct}
\end{align}
where
\begin{align*}
Q(l,m) = (-1)^{(m+|m|)/2} \sqrt{\frac{(2l+1)(l+|m|)!}{2(l-|m|)!}},
\end{align*}
$\left(D^l_{m' m}(\hat{R})\right)^\ast$ is the Wigner $D$-matrix for rotating the molecular frame wave function to the laboratory frame with Euler angles $\hat{R}$,
and $C_{lm'}$ is the structure parameter~\cite{zhao2010,  zhao2011}.
The factors $B_{im}$ have different expressions for $F>0$ and $F<0$ due to the asymptotic expression of the spherical harmonics at opposite field directions.
Notably,  for the case of ionizing two orbitals with the same magnetic quantum number $m$ of an atom,  Eq.~\eqref{eq:motif} and \eqref{eq:gamma} coincide with the model by Pabst \textit{et al.}~\cite{pabst2016},
where the parity of the orbital is given by $(-1)^l$.

Eq.~\eqref{eq:gamma} possesses several desirable features for modeling multiorbital TI:

I.) It is described at the same level of accuracy as the MO-ADK theory.

II.) It is computationally efficient.

III.) It is applicable to any molecular systems once the structure parameters for the MO-ADK theory are determined.

IV.) It is less likely to overestimate the coherence compared to the model in Eq.~\eqref{eq:others_model}.

The physical interpretations of Eqs.~\eqref{eq:motif} and \eqref{eq:gamma} are quite intuitive.
Consider a scenario where TI occurs only at every half laser cycle,  causing population of the $i$th and $j$th states to build up. 
The resulting coherence will exhibit constructive or destructive interference depending on the energy difference $E_i - E_j$ and structures of the orbitals involved. 
In general,  the coherence is anisotropic.
For instance,  in a linear molecule,  if the $i$th and the $j$th states are formed by ionizing a $\sigma$ and a $\pi$ orbital,  respectively,  their coherence will be zero when the molecule is aligned with the laser polarization, despite the non-zero ionization rates.
This is because ionized electrons from a $\sigma$ or a $\pi$ orbital have distinct momentum distributions when the molecule is aligned.
Only at certain alignment angles,  where the ionized electrons from the two different orbitals share similar momentum distributions,  will their coherence strengthen.
A special case arises when the $i$th and $j$th states are spin-orbit states formed from ionizing the same orbital.
In this situation,  the ionized electrons will possess highly similar momentum distributions, leading to constructive interference of their coherence throughout the laser pulse,  resulting in substantial coherence after the pulse ends.
Coherence between spin-orbit states has been studied experimentally in heavy noble gas atoms~\cite{goulielmakis2010} and halogen-containing molecules~\cite{kobayashi2020,  kobayashi2020a}.

\subsection{The DM-SDI model}

In the following,  we will incorporate Eqs.~\eqref{eq:motif} and~\eqref{eq:gamma} into the DM-SDI model developed recently by our group~\cite{yuen2022, yuen2023}.
An overview of the DM-SDI model is provided,  while additional details can be found in the references mentioned.
The equations of motion for the density matrices $\rho^{(q)}$ are given by
\begin{align}
 \frac{d}{dt}\rho^{(q)}(t) = -\frac{i}{\hbar}[H^{(q)}(t),\rho^{(q)}(t)] + \Gamma^{(q)}(t),
 \label{eq:EOM}
\end{align}
where $q$ is the charge of the molecule and $H^{(q)}$ is the Hamiltonian with the laser coupling term $-\vec{d}\cdot \vec{E}$.

The ionization rate matrix for the neutral state is described by
\begin{align*}
\Gamma^{(0)}(t) = -\sum_i \rho^{(0)}(t) W^{(0)}_i,
\end{align*}
where $W^{(0)}_i$ is the ionization rate from the neutral to the $i$th ionic state.

Motivated by Eq.~\eqref{eq:motif},  the ionization rate matrix for the ionic state $\Gamma^{(1)}(t)$ is modeled as
\begin{align}
\Gamma^{(1)}_{ij}(t) &= \rho^{(0)}(t) \sum_m \gamma^{(0)}_{im} (t) \left(\gamma^{(0)}_{jm}(t)\right)^\ast  \nonumber \\ 
&- \rho^{(1)}_{ij}(t) \sqrt{\sum_n W^{(1)}_{n \leftarrow i}(t)} \sqrt{\sum_n W^{(1)}_{n \leftarrow j}(t)},
\label{eq:gamma1}
\end{align}
where $\gamma^{(0)}_{im}$ is the partial ionization amplitude (as defined in Eq.~\eqref{eq:gamma}) from the neutral to the $i$th ionic state,  and $W^{(1)}_{n \leftarrow i}$ corresponds to the ionization rate from the $i$th ionic state to the $n$th state of the dication.
The diagonal elements describe the population transfer from TI of the neutral and TI to the dication. 
The off-diagonal elements of the first term describes the build up of coherence from TI,  analogous to Eq.~\eqref{eq:motif},  while the second term describes dephasing due to population loss from TI.
For the model without the coherence from TI,  the first term in Eq.~\eqref{eq:gamma1} is replaced by $\delta_{ij}  \rho^{(0)}(t)W^{(0)}_i(t)$,  as presented in our previous work~\cite{yuen2022,  yuen2023}.

Finally,  the ionization matrix for the dication $\Gamma^{(2)}$ is modeled similarly as
\begin{align}
\Gamma^{(2)}_{mn}(t) = \sum_i \rho^{(1)}_{ii}(t) \sum_\mu \gamma^{(1)}_{m \leftarrow i, \mu} (t) \left(\gamma^{(1)}_{n \leftarrow i,  \mu }(t)\right)^\ast,
\label{eq:gamma2}
\end{align}
where $\gamma^{(1)}_{m \leftarrow i, \mu}$ is the partial ionization amplitude for TI from the $i$th ionic state to the $m$th state of the dication.
For the model without TI coherence,  only the diagonal terms are retained,  resulting in $\Gamma^{(2)}_{mn}(t) = \delta_{mn} \sum_i \rho^{(1)}_{ii}(t) W^{(1)}_{n \leftarrow i}(t)$~\cite{yuen2022,  yuen2023}.

Upon solving Eq.~\eqref{eq:EOM} for each charge state,  the populations of different dication states,  $ \rho^{(2)}(\theta,  t\to \infty)$,  are obtained after the laser pulse for various alignment angle $\theta$ between the molecular axis and the laser polarization direction.
The populations are then angular averaged to account for the orientation of the molecule.
For each dication state,  the averaged yields are then mapped to the kinetic energy release (KER) and branching ratio.
\subsection{Application to N$_2$ and O$_2$}

To provide a foundation for further discussions, we present a brief overview of the mechanism of SDI of N$_2$ and O$_2$. 
For more details,  please refer to Ref.~\cite{yuen2022} for N$_2$ and Ref.~\cite{yuen2023} for O$_2$.

In the case of N$_2$,  the first step in SDI involves ionization of the $3\sigma_g$(15.6 eV),  $1\pi_{u\pm}$(16.9 eV),  and $2\sigma_u$ (18.8 eV) orbitals to form the $X^2\Sigma_g^+,  A^2\Pi_{u\pm}$,  and $B^2\Sigma_u^+$ states of N$_2^+$,  respectively. 
For brevity hereafter,  these states will be referred as the $X$,  $A_\pm$,  and $B$ states of N$_2^+$.
The laser field couples the $X-A_\pm$ states with dipole moment $d_x = \mp 0.17 \, ea_0$ and the $X-B$ states with dipole moment $d_z = 0.75 \, ea_0$.
Subsequently,  an electron from the $3\sigma_g,  1\pi_u$,  and $2\sigma_u$ orbitals of these ionic states are tunnel ionized to form the dication states.
Note that the dication states with at least one hole in the highest occupied molecular orbital,  i.e. $3\sigma_g$,  are considered non-dissociative.

In the case of O$_2$,  the first step in SDI involves ionization of the $1\pi_{g\pm}$(12.3 eV),  $1\pi_{u\pm}$(16.7 eV),  and the $3\sigma_g$(18.2 eV) orbitals to form the  $X^2\Pi_{g\mp}$,  $a^4\Pi_{u\pm}$,  and $b^4\Sigma_g^-$ states of O$_2^+$.
For brevity hereafter,  these states will be referred as the $X_\mp$,  $a_\pm$,  and $b$ states of O$_2^+$.
The $a_\pm$ and $b$ states are then coupled by the laser field with a dipole moment of $d_x = \mp 0.205 \, ea_0$.
The dication states are then formed through the second TI from the $1\pi_{g},  1\pi_{u}$,  and  $3\sigma_g$ orbitals of these intermediate states.
A key distinction in the SDI dynamics of O$_2$ compared to N$_2$ is the presence of laser couplings between certain dication states,  allowing for population transfer between them.

It is important to note that when modeling the coherence of ionic states, degenerate states must be treated separately. 
In our previous approach~\cite{yuen2022, yuen2023},  the $\Pi$ state was represented only by the $\Pi_x$ state,  as the $\Pi_y$ state has zero dipole moment in the $xz$ polarization plane. 
Additionally,  to simulate the formation of both the $\Pi_x$ and $\Pi_y$ states,  the ionization rate for the $\Pi_x$ state was set to be twice as fast.
While this approach yields correct results when considering either TI or laser coupling alone,  it leads to approximately 20\% differences in the population of ions and dications when both TI and laser couplings are considered,  compared to the approach that treats the $\Pi_x$ and $\Pi_y$ states separately. 
This discrepancy arises because,  although the $\Pi_y$ state is formed at the same rate as the $\Pi_x$ state,  it does not contribute to the evolution of coherence due to the absence of laser coupling. 
Consequently,  assigning the $\Pi_y$ state to be the $\Pi_x$ state causes the evolution of ionic coherence to differ slightly,  resulting in a slight variation in the final populations of ions and dications.

\section{Influence of TI coherence in single ionization~\label{sec3}}
After undergoing tunnel ionization, different ionic states can couple to each other through the laser couplings, resulting in an interplay between their population and coherence dynamics. 
Therefore, it is important to investigate the influence of coherence from TI on the overall population and coherence of the system.

In our study,  we consider a 6 fs linearly polarized Gaussian laser pulse with a wavelength of 800 nm.
The peak intensity of the pulse is set to 3 $\times$ 10$^{14}$ W/cm$^2$,  such that the TI to the dication states can be safely neglected.
To examine the impact of TI coherence,  we solve Eq.~\eqref{eq:EOM} with and without the off-diagonal terms in Eq.~\eqref{eq:gamma1} for N$_2$ and O$_2$ at each alignment angle $\theta$, and obtain the ionic density matrix $\rho^{(1)}(\theta,  t_f)$ after the pulse ends.
Note that we assume the molecule is randomly oriented in space for our analysis.

\begin{figure}[t]
\includegraphics[scale=0.5]{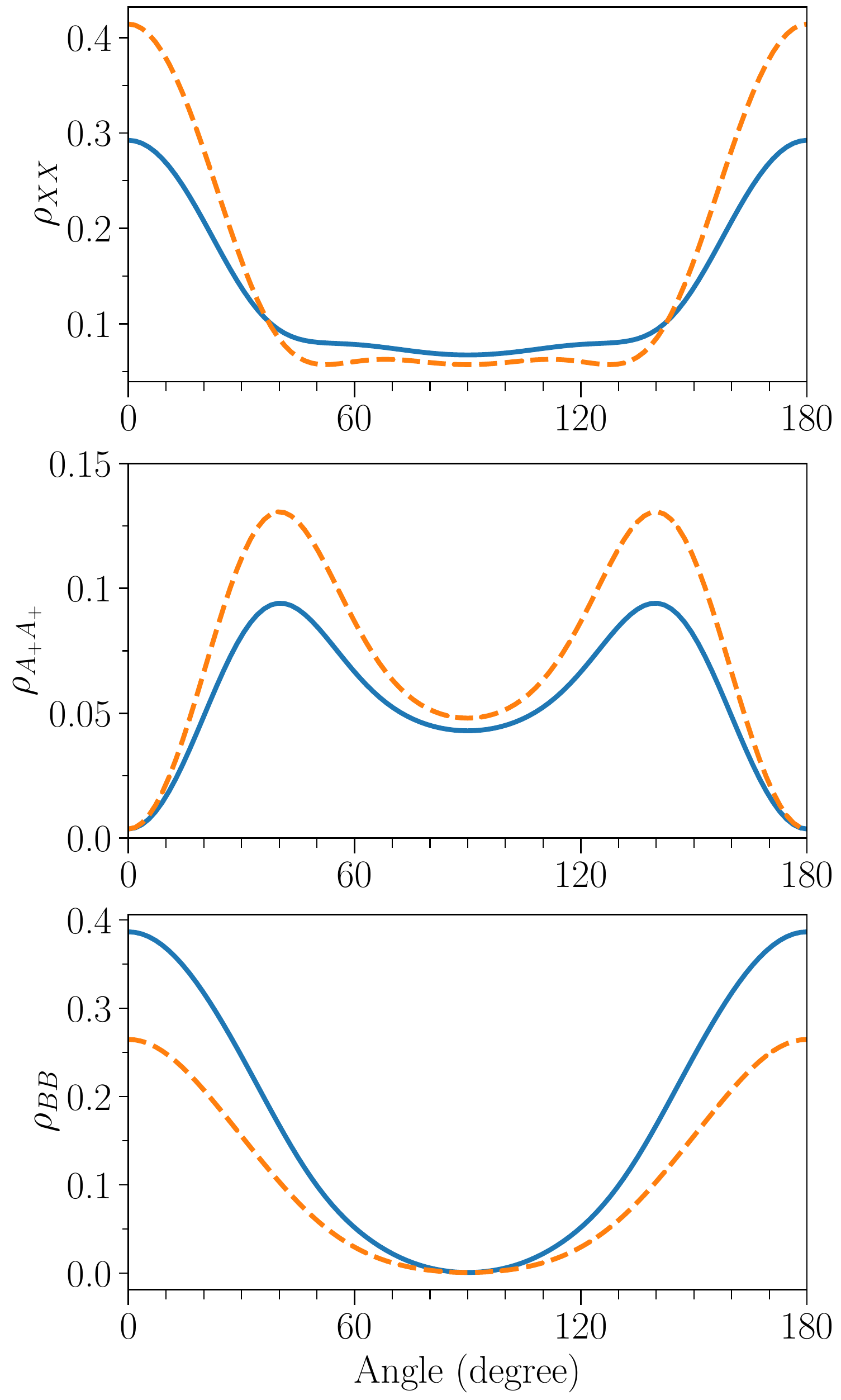} 
\caption{Population of the $X^2\Sigma_g^+$ (top),  $A^2\Pi_{u+}$ (middle),  and $B^2\Sigma_u^+$ (bottom) states of N$_2^+$ after the laser pulse at different alignment angles.  Solid lines represent the results considering TI coherence, while dashed lines represent the results without TI coherence.
The laser used in the calculation is linearly polarized with a peak intensity of 3 $\times 10^{14}$ W/cm$^2$, a pulse duration of 6 fs, and a wavelength of 800 nm.}
\label{fig:N2_pop1}
\end{figure}

\begin{figure}[t]
\includegraphics[scale=0.55]{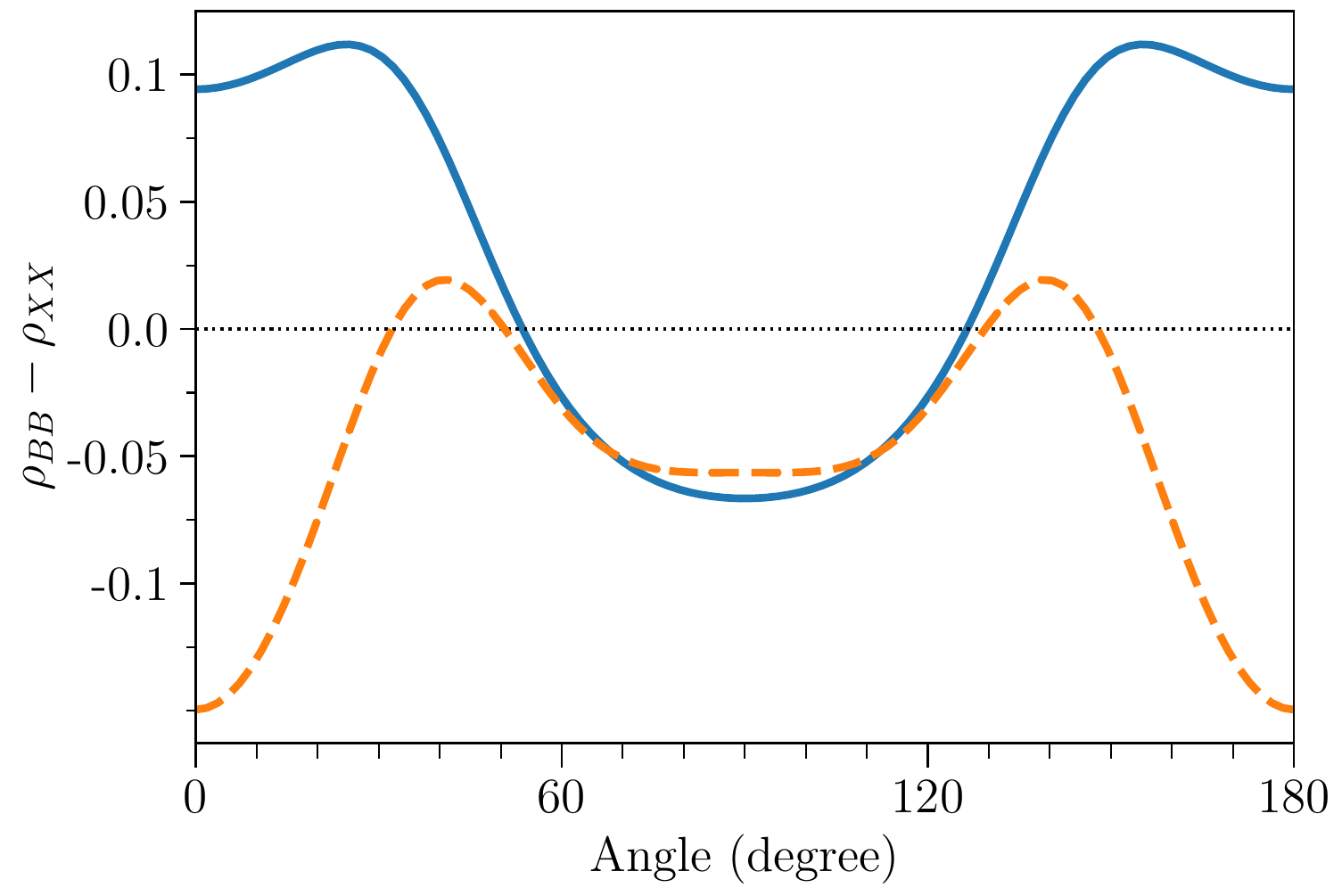} 
\caption{Same as Fig.~\ref{fig:N2_pop1},  but for the difference between the population of the $B^2\Sigma_u^+$ and the $X^2\Sigma_g^+$ states of N$_2^+$.}
\label{fig:N2_pop_diff}
\end{figure}

To assess the overall influence of TI coherence on the ionic population of N$_2^+$, we present the ionic populations as a function of the alignment angle in Fig.~\ref{fig:N2_pop1}. 
It can be observed that,  with the inclusion of TI coherence,  the populations of the $A_+$ and $B$ states are generally smaller and larger compared to the results without TI coherence,  respectively. 
When averaged over alignment angles,  with the TI coherence,  the average population of the $A$ and $B$ state is about 30\% lower and 37\% higher.
The average population of the $X$ state for both models are approximately the same.

An important observation is that when TI coherence is included,  the population of the $B$ state surpasses that of the $X$ state over a much wider range of angles compared to the results without TI coherence (see Fig.~\ref{fig:N2_pop_diff}).
We also observe that,  with the TI coherence,  the average population of the $A$ state is around $0.13$,  which exceeds the average populations of the $X$ and $B$ states ($\sim 0. 1$ for both).
Assuming the molecule orients in space randomly,  a narrow range of angles for the population inversion would make air lasing less likely to occur. 
Our results with TI coherence then align with the mechanism proposed by Xu \textit{et al.}~\cite{xu2015},  suggesting that the $A$ state acts as a population reservoir for the $X$ state,  enabling population inversion for the $B$ state. 
Therefore,  our findings indicate that TI coherence plays a crucial role in the post-ionization dynamics of N$_2^+$.

\begin{figure}[t]
\includegraphics[scale=0.6]{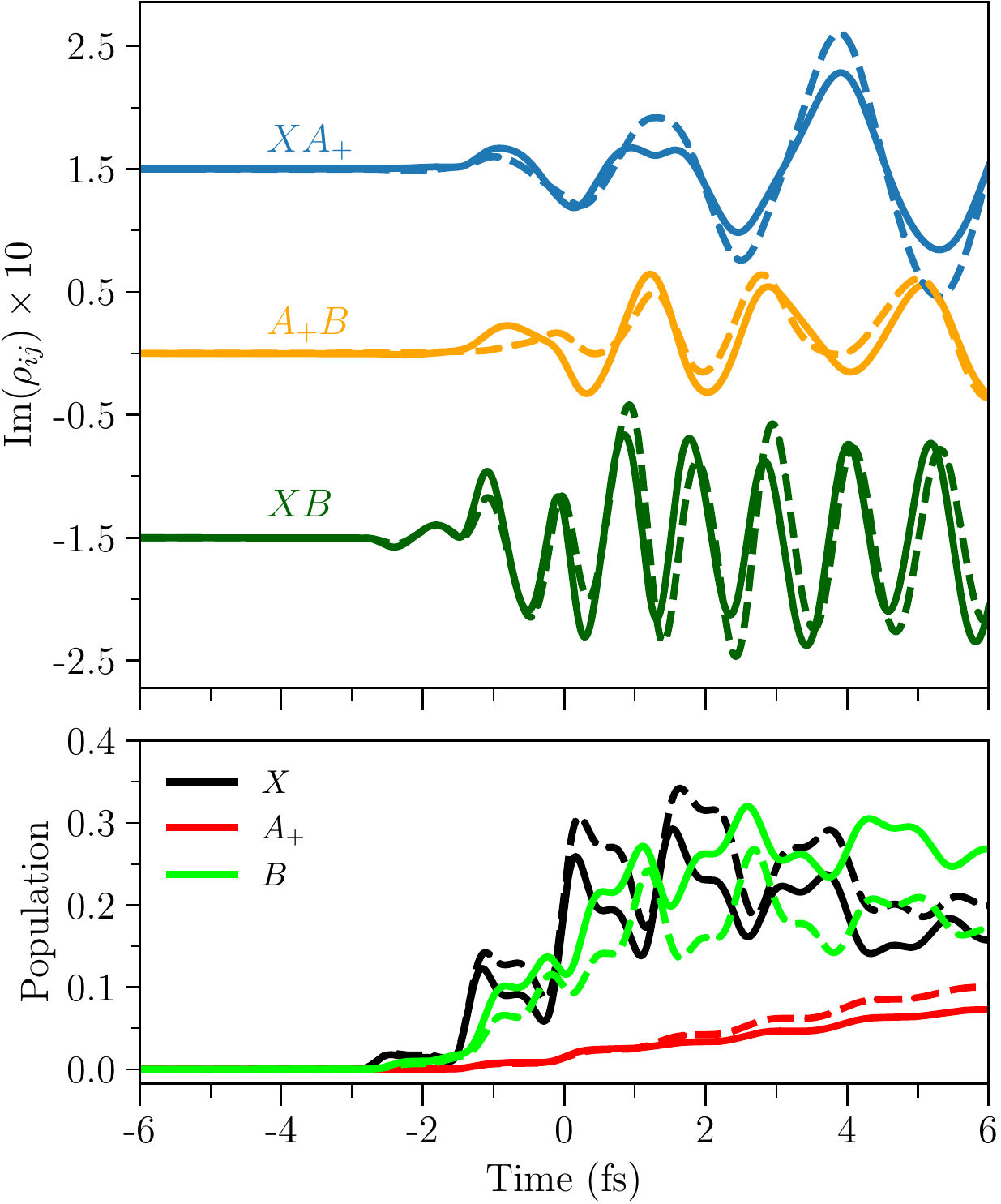} 
\caption{Evolution of the $X^2\Sigma_g^+ - A^2\Pi_{u+}$,  $A^2\Pi_{u+} - B^2\Sigma_u^+$,  and $X^2\Sigma_g^+ - B^2\Sigma_u^+$ coherences (top) of N$_2^+$ and the respective population (bottom) at $\theta = 27^{\circ}$.  Solid lines represent results with TI coherence, while dashed lines depict results without TI coherence.  The same laser parameters as in Fig.~\ref{fig:N2_pop1} are used in the calculations. To enhance clarity, the density matrix elements are vertically shifted for better visualization}
\label{fig:weak_N2_27}
\end{figure}

To further elucidate the role of TI coherence on the post-ionization dynamics, we display the evolution of ionic population and coherence of N$_2^+$ at $\theta=27^\circ$ in Fig.~\ref{fig:weak_N2_27}.
Note that only the imaginary part of the off-diagonal elements are plotted,  since the evolution of ionic population, 
\begin{align}
\frac{d}{dt}\rho^{(1)}_{ii}(t)  = -2 \sum_{l} \vec{d}_{il}\cdot\vec{E}(t)\, \mathrm{Im}\left(\rho^{(1)}_{li}(t)\right) + \Gamma^{(1)}_{ii}(t),
\label{eq:ion_pop}
\end{align}
depends solely on their imaginary part.
We observe that the $X-A_+$, $A_+ - B$, and $X - B$ coherence exhibit similar qualitative behavior with or without TI coherence. 
This suggests that the qualitative behavior of the ionic coherence is primarily determined by the laser couplings at this angle.
However, when TI coherence is considered,  the population of the $B$ state after the pulse is 36\% higher compared to the results without TI coherence. 
This increase is attributed to the fact that the $X-B$ coherence is 40\% stronger when TI coherence is included at $t \sim 0$ fs,  leading to a surge in the $B$ population.
Conversely, with TI coherence, the population of the $A_+$ state after the pulse is 38\% lower than the results without TI coherence. 
This reduction is a consequence of the $X- A_+$ coherence being approximately 40\% weaker at $t \sim 4$ fs when TI coherence is considered. 
Consequently, fewer $A_+$ states are formed via laser coupling.
Hence,  we see that TI coherence influences the ionic population of N$_2^+$ by quantitatively altering the $X-A_+$ and $X-B$ coherences.

\begin{figure}[t]
\includegraphics[scale=0.5]{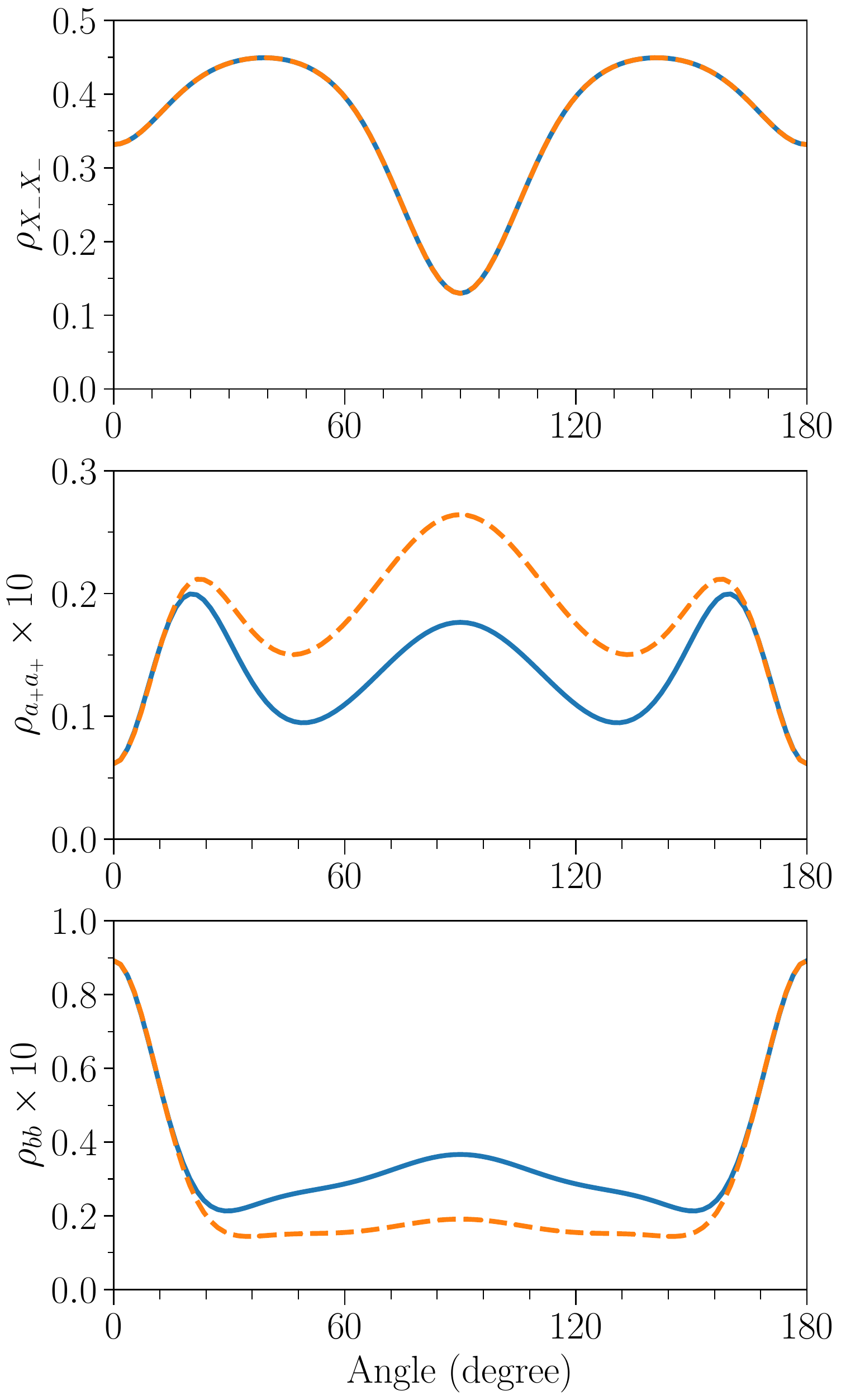} 
\caption{Same as Fig.~\ref{fig:N2_pop1},  but for the $X^2\Pi_{g-}$ (top),  $a^4\Pi_{u+}$ (middle),  and $b^4\Sigma_g^-$ (bottom) states of O$_2^+$. }
\label{fig:O2_pop1}
\end{figure}

In Fig.~\ref{fig:O2_pop1}, we present the ionic populations of O$_2^+$ as a function of the alignment angle. 
Similar to the N$_2^+$ case, the populations of the $a_\pm$ and $b$ states of O$_2^+$ are influenced by the presence of TI coherence.
Since only the $a_\pm$ and $b$ states of O$2^+$ are coupled by the laser, the population of the $X$ state remains unchanged with or without TI coherence. 
However, when TI coherence is considered,  the angular-averaged population of the $b$($a_+$) state is enhanced (suppressed) by about 39\%. 
Therefore, similar to the case of N$_2^+$, the presence of TI coherence alters the ionic population of O$2^+$ by roughly 40\%.

\begin{figure}[t]
\includegraphics[scale=0.6]{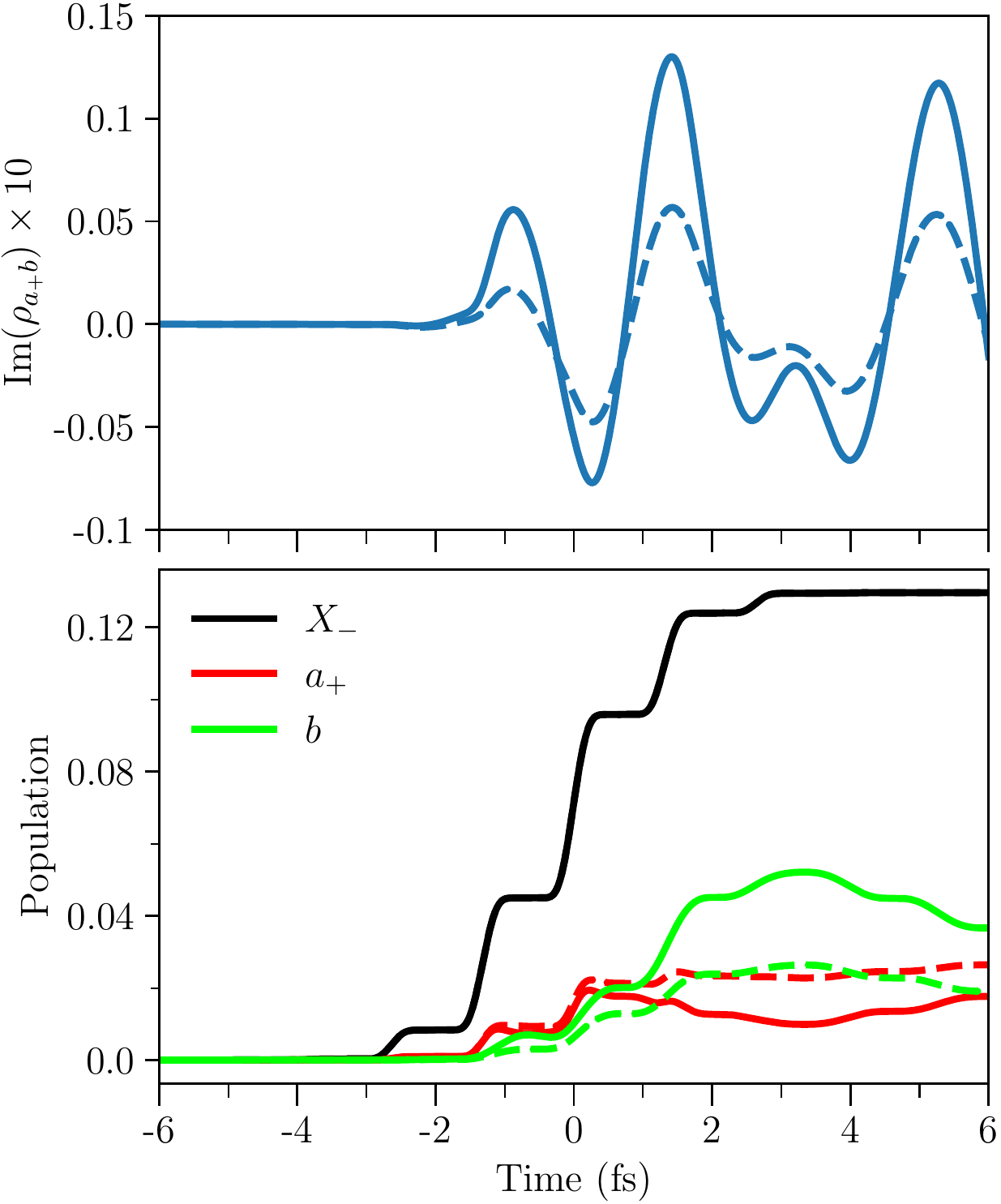} 
\caption{Same as in Fig. ~\ref{fig:weak_N2_27}, but for evolution of the $a^4\Pi_{u+}-b^4\Sigma_g^-$ coherence (top) of O$_2^+$ and the respective population (bottom) at $\theta = 90^{\circ}$.  }
\label{fig:weak_O2_90}
\end{figure}

The role of TI coherence in the case of O$_2^+$ exhibits similarities to that of N$_2^+$. 
This similarity is demonstrated in Fig.~\ref{fig:weak_O2_90}, which illustrates the evolution of the $a_+-b$ coherence and the ionic population of O$_2^+$ at an alignment angle of $\theta = 90^{\circ}$.
At around $t \sim 1$ fs, the $a_+-b$ coherence is more than twice as strong when the TI coherence is taken into account. 
Consequently,  this increased coherence leads to a significant rise in the population of the $b$ state.

An important question arises as to whether the effect of TI coherence in single ionization can be observed in experiments. 
For the case of N$_2$,  Kleine \textit{et al.}~\cite{kleine2022} conducted an attosecond transient absorption spectroscopy (ATAS) experiment using the N K-edge to determine the relative population of the $X$, $A$, and $B$ states of N$_2^+$ after multiorbital TI using an 800 nm,  50 fs laser pulse with a peak intensity of $4.5 \times 10^{14}$ W/cm$^2$. 
From their measurement,  they estimated the time-dependent electronic state population of N$_2^+$ and observed that the $X$ and $B$ states have nearly equal populations, while the population of $A$ is low. 
However, it is worth noting that the pulse duration of the IR field employed in their study is considerably longer than the one considered in this work. 
It would be highly valuable to perform similar experiments using a few-cycle IR pulse to validate our findings. 
Meanwhile,  efforts are underway to simulate an ATAS experiment based on the current model.


For the case of O$_2$,  table-top ATAS experiments will be challenging to perform since one has to generate the x-ray for the O K-edge.
However, alternative experiments based on the strong field dissociation of O$2^+$ induced by an IR field have been available~\cite{de2011}. 
In these experiments, dissociation occurs through laser coupling between the $a^4\Pi_{u}$ and the dissociative $f^4\Pi_{g}$ states. 
A previous theoretical study~\cite{xue2018} has demonstrated that including the laser coupling between the $a^4\Pi_{u}$ and $b^4\Sigma_g^-$ states significantly enhances the agreement on the quantum beat spectrum with experiments.
Furthermore, Xue \textit{et al.}~\cite{xue2021} revealed that incorporating vibronic coherence in the reduced density matrix of O$_2^+$ can introduce distinct features in the quantum beat spectrum. 
Thus, we anticipate future simulations utilizing our coherence model and experiments investigating the strong field dissociation of O$_2^+$.

\section{Influence of TI coherence in sequential double ionization~\label{sec4}}
As the peak laser intensity increases, there is a possibility for the nascent ionic states to undergo tunnel ionization again, resulting in the formation of dications. 
Building upon the demonstration in the previous section, the presence of TI coherence would enhance the population of certain intermediate ionic states, thereby increasing the yield of specific dication states. 
Notably, in the case of O$_2^{2+}$, laser couplings between dication states are known to play an important role in the post-ionization dynamics~\cite{yuen2023}. 
Therefore, it becomes necessary to consider TI coherence between dication states, which can be described by Eq.~\eqref{eq:gamma2}.
As mentioned in Sec.~\ref{sec2}.B, the solutions to Eq.\eqref{eq:EOM} are used to obtain the dication yields, which are subsequently averaged over alignment angles. 
These averaged yields are then mapped to the KER of each state. 
In this section, we compare the KER spectra and angle-dependent dication yields obtained with and without TI coherence for both N$_2$ and O$_2$ at a peak intensity of 1.2 $\times 10^{15}$ W/cm$^2$. 
The remaining laser parameters are identical to those described in Sec.~\ref{sec3}.

\begin{figure}[t]
\includegraphics[scale=0.45]{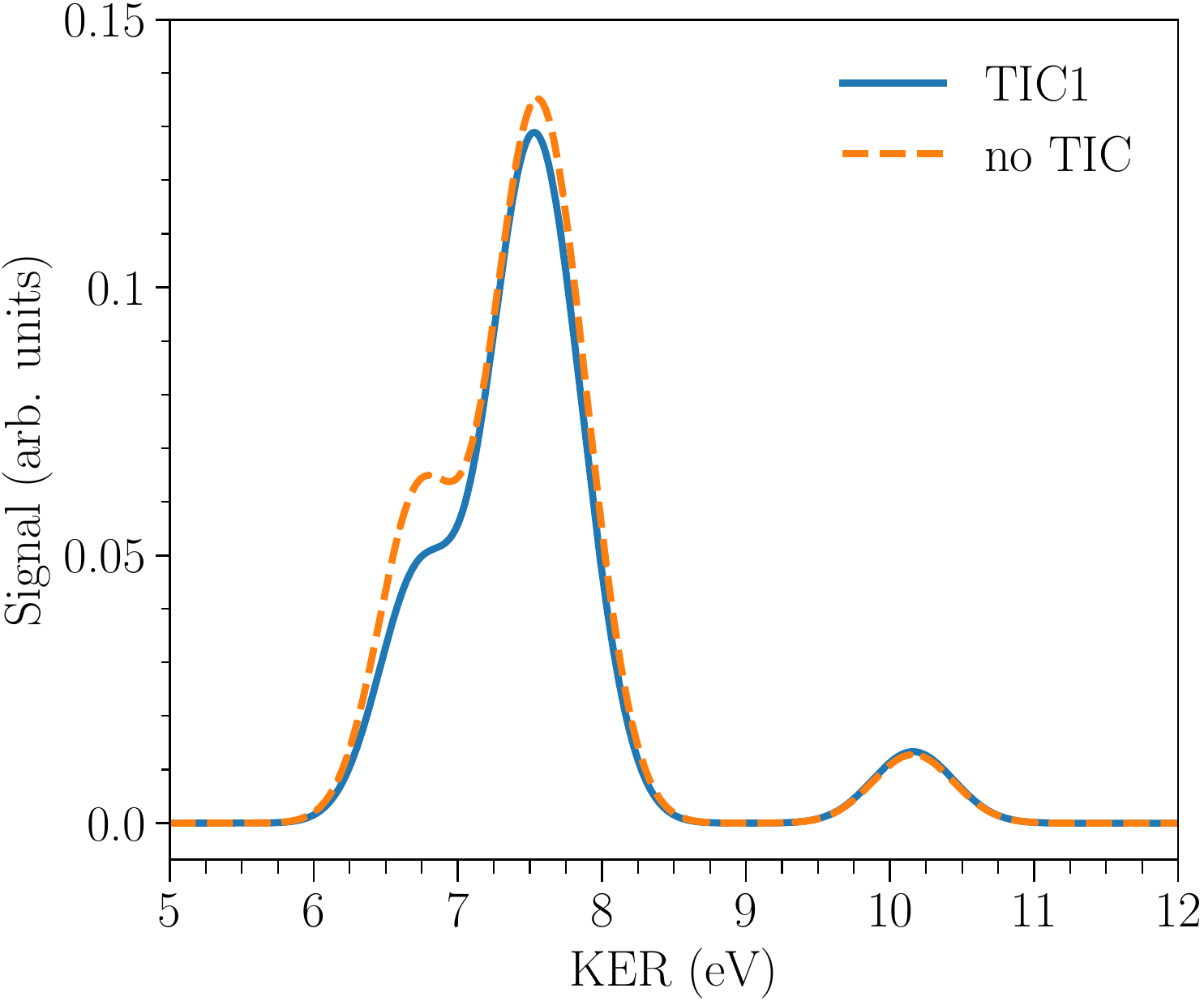} 
\caption{Comparison of KER spectra for $\rm N^+ + N^+$ with and without TI coherence (TIC1/no TIC).
The laser parameters are identical to those shown in Fig.~\ref{fig:N2_pop1}, but with a peak intensity of 1.2 $\times 10^{15}$ W/cm$^2$.}
\label{fig:N2_KER}
\end{figure}

Figure~\ref{fig:N2_KER} illustrates the KER spectra for $\rm N^+ + N^+$ with randomly oriented N$_2$.
Overall, the inclusion of TI coherence between the ionic states leads to minor quantitative changes in the KER spectrum. 
Specifically,  we observe that the KER peaks at approximately 6.8 eV and 7.5 eV are reduced by approximately 30\% and 5\% when TI coherence is included.
The 6.8 eV peak corresponds to a state with two holes in the $1\pi_u$ orbitals, which can only be formed through the $A$ state of N$_2^+$. 
As discussed in the previous section, the population of the $A$ state is suppressed when TI coherence is included,  such that there is a reduction of the 6.8 eV peak.
Regarding the 7.5 eV peak, it comprises states with configurations $(1\pi_u)^{-2}$ and $(2\sigma_u^{-1})(1\pi_u)^{-1}$,  which can be formed through both the $A$ and $B$ states of N$_2^+$.
Since the population of the $B$ state is enhanced when TI coherence is considered,  the suppression of the 7.5 eV peak is weaker.
When considering the combined effects, the ratio of the 6.8 to 7.5 eV peaks is lowered by only about 20\% when TI coherence is included,  which falls within typical experimental uncertainties. 
Hence,  identifying the influence of TI coherence in SDI of N$_2$ through the KER spectrum is unlikely in experiments~\cite{voss2004, wu2011}.

\begin{figure}[t]
\includegraphics[scale=0.45]{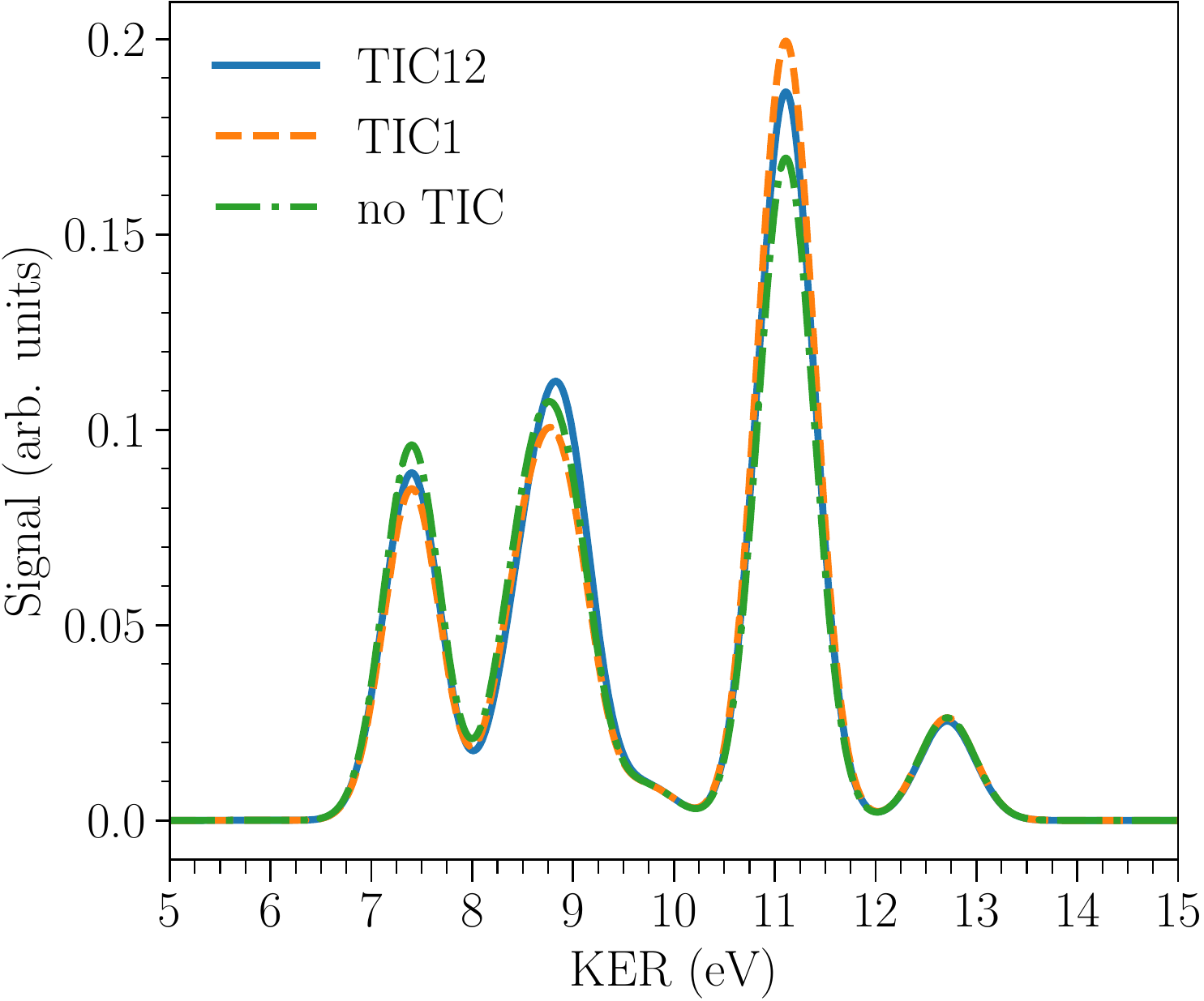} 
\caption{KER spectra for $\rm O^+ + O^+$ for three different cases: TIC12, which represents results with TI coherence in the first and second TI; TIC1, which represents results with TI coherence only in the first TI; and no TIC,  which represents results without TI coherence in both the first and second TI.}
\label{fig:O2_KER}
\end{figure}

Figure~\ref{fig:O2_KER} displays the KER spectra for $\rm O^+ + O^+$ with randomly oriented O$_2$.
Three cases are considered in the SDI of O$_2$: TIC12,  with the inclusion of TI coherence in the first and second TI; TIC1,  with the inclusion of the TI coherence in the first TI only; and no TIC,  no TI coherence in either TI.
The KER spectra exhibit similar qualitative behavior in all three cases, with some small quantitative differences. 
In the TIC1 case,  the 11.1 eV peak is approximately 15\% higher compared to the no TIC case,  while the 7.4 and 8.8 eV peaks in the TIC1 case are about 13\% lower than the no TIC case.
As discussed in the previous section, the TI coherence of the ion enhances the formation of the $b$ state of O$_2^+$, resulting in the increased height of the 11.1 eV peak,  which corresponds to states that can be formed via the $X$ and $b$ state.
Conversely, the TI coherence suppresses the formation of the $a$ state,  leading to lower peaks at 7.4 and 8.8 eV, which corresponds to states that can be formed via the $X$ and $a$ states.
When the TI coherence of the dication is considered (TIC12), the height of the 11.1 eV peak decreases by approximately 7\% compared to the TIC1 case,  as enhanced laser couplings transfer more population from the state corresponds to the 11.1 eV peak to the states responsible for the 7.4 and 8.8 eV peaks.
The peak ratios between 7.4 ,  8.8,  and 11.1 eV peaks are 0.56:0.63:1,  0.42:0.5:1,  and 0.47:0.60:1 for the case of no TIC,  TIC1,  and TIC12,  respectively.
Similarly to the case of N$_2$,  the effects of TI coherence are unlikely to be identified in the experimental KER spectrum for $\rm O^+ + O^+$ due to the small quantitative changes compared to experimental uncertainties~\cite{voss2004,wu2011}.

Based on the N$_2$ and O$_2$ cases, it can be concluded that when comparing angular-averaged quantities such as KER spectra, modeling SDI without considering the coherence from TI is sufficiently accurate, thus validating our previous approach~\cite{yuen2022, yuen2023}. 
We also observe that the quantitative differences caused by the inclusion of TI coherence are generally smaller in the case of SDI than in the case of single ionization.
This is because the TI rates from ions to dications are typically faster than the population transfer rates induced by laser couplings in SDI,  thereby diminishing the importance of TI coherence. 
Although the effects of TI coherence may not be evident in the KER spectra, the angular-dependent yield of dication states is expected to be more sensitive to TI coherence since laser coupling would be preferentially enhanced at certain alignment angles.
It is important to note that while the angular distribution of different KER peaks can be directly measured using coincidence imaging, the measured distribution is unlikely to match the calculated results due to the post-ionization alignment effect of the ionic fragments~\cite{voss2004, tong2005}. 
Instead, the angle-dependent yield should be extracted using the rotational wave packet of the molecule, as demonstrated by Lam \textit{et al.}~\cite{lam2020} for CO$_2$.

\begin{figure}[t]
\includegraphics[scale=0.5]{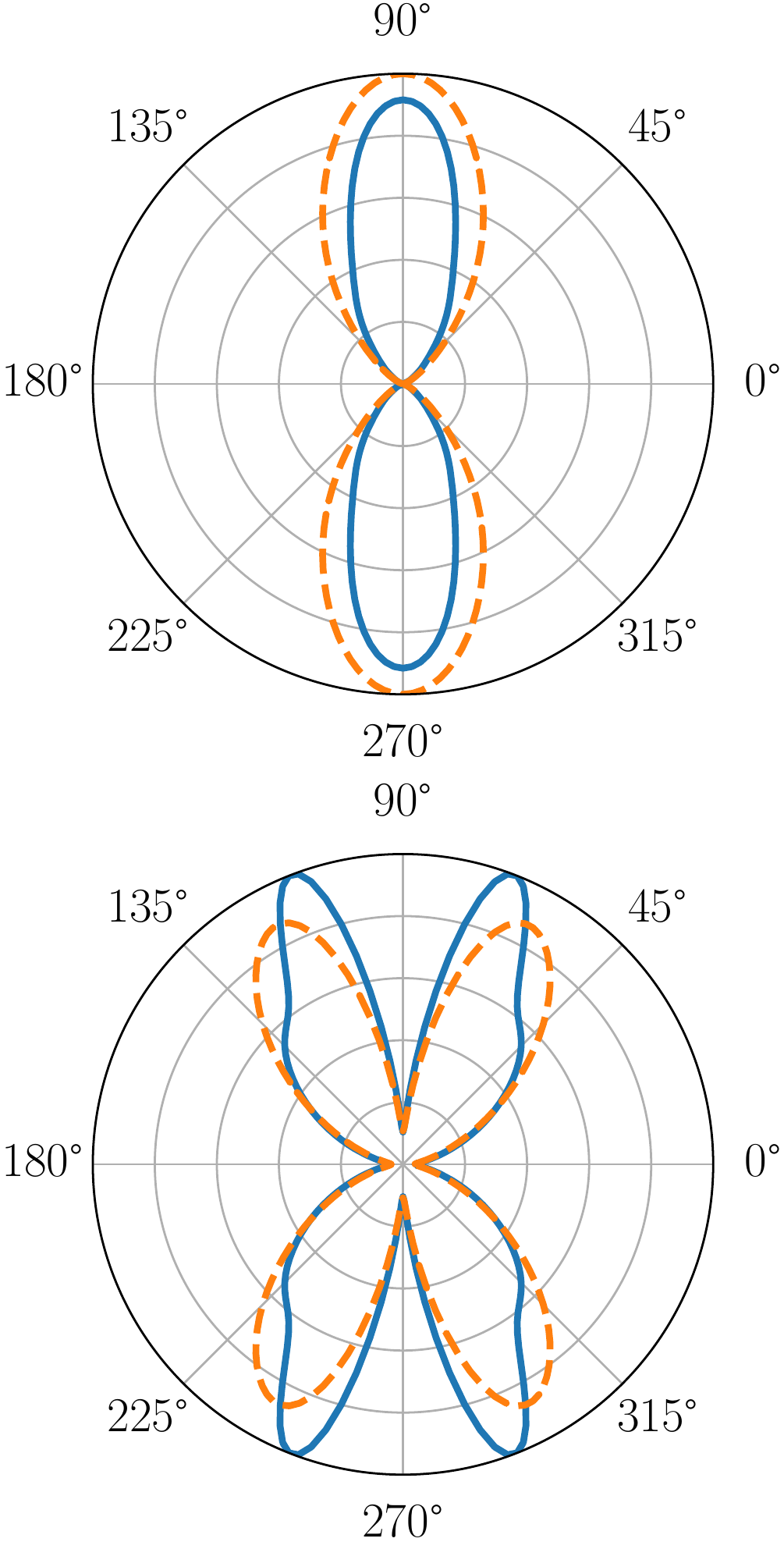} 
\caption{Angle-dependent yield of states responsible for the 6.6 eV (top) and 10.1 eV (bottom) KER peaks for N$_2$.
Solid or dashed lines represent results with or without TI coherence.}
\label{fig:N2pp_ang_yield}
\end{figure}

Figure~\ref{fig:N2pp_ang_yield} displays the calculated angular distributions for the 6.6 and 10.1 eV KER peaks for N$_2$  with and without TI coherence. 
These peaks are chosen because they correspond to the formation of states with electronic configuration $1\pi_u^{-2}$ and $2\sigma_u^{-1} 1\pi_u^{-1}$,  respectively.
We see that the angular distribution of the 6.6 eV peak appears similar with and without TI coherence. 
However, for the 10.1 eV peak, the distribution changes from a clover-like shape to a butterfly-like shape when the TI coherence is considered.

\begin{figure}[t]
\includegraphics[scale=0.5]{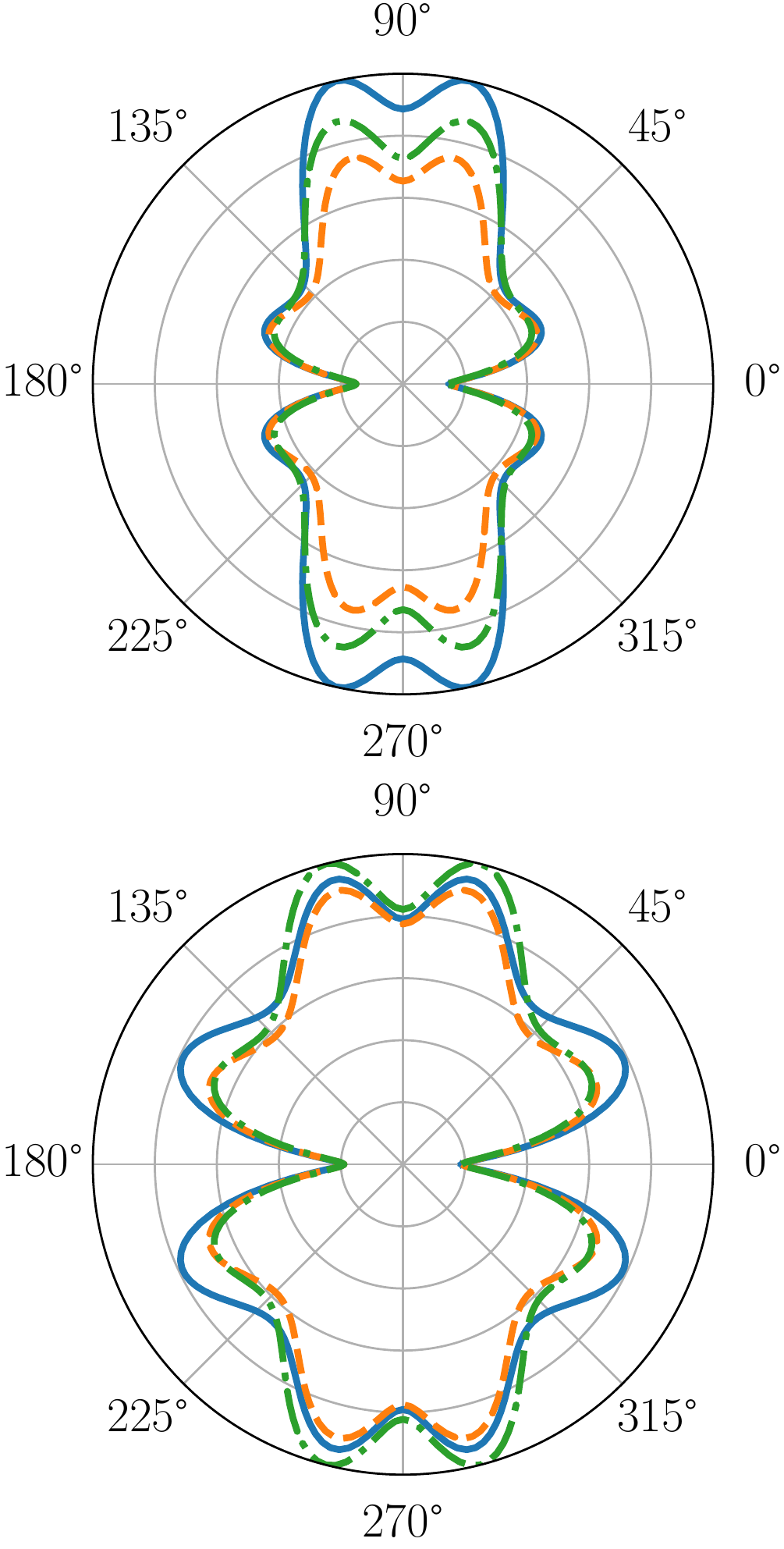} 
\caption{Angle-dependent yield of states responsible for the 7.4 eV (top) and the 8.8 eV (bottom) KER peaks for O$_2$ from the three models in Fig.~\ref{fig:O2_KER}.}
\label{fig:O2pp_ang_yield}
\end{figure}

In Fig.~\ref{fig:O2pp_ang_yield},  we present the calculated angular distributions for the 7.4 and 8.8 eV KER peaks for O$_2$ for the cases of TIC12, TIC1, and no TIC. 
These peaks are selected due to the more visible changes in their distributions among the three cases. 
Although both peaks correspond to states with the configuration $1\pi_u^{-1}1\pi_g^{-1}$, their angular distributions differ due to laser couplings.
The angular distributions of both peaks are found to be similar for the TIC1 and no TIC cases. 
However, in the TIC12 case, the distribution for the 7.4 eV peak appears longer, while the distribution for the 8.8 eV peak appears wider compared to the other two cases.

Based on these observations, it can be concluded that the angle-dependent yield serves as a more suitable quantity to verify the TI coherence model. 
Therefore, we anticipate future pump-probe experiments, similar to the work presented in Ref.~\cite{lam2020}, that can extract the angle-dependent yield of N$_2$ and O$_2$. 
In these experiments, the pump pulse can impulsively align the molecules, while the intense few-cycle probe pulse can trigger the SDI of N$_2$ or O$_2$. 
By extracting the angle-dependent yield, these experiments can provide a direct test of the TI coherence model.
Furthermore, in addition to the angle-dependent yield, such experiments can also verify the alignment dependence of the KER spectra as described in our previous works~\cite{yuen2022,yuen2023}. 
This verification would contribute to further validating the accuracy of the DM-SDI model and its predictions.

\section{Summary and Outlook~\label{sec5}}
In summary, we have proposed and implemented a coherence model for multiorbital TI in the framework of the DM-SDI model. 
The coherence model is a natural extension of the MO-ADK theory and does not require additional parameters. 
By incorporating the TI coherence and dephasing effects in Eqs.~\eqref{eq:gamma1} and ~\eqref{eq:gamma2}, we have completed the DM-SDI model while maintaining computational efficiency.

We have applied this new model to investigate single ionization and SDI of N$_2$ and O$_2$. 
In the case of single ionization, we found that TI coherence leads to changes of approximately 40\% in the angular-averaged ionic populations of N$_2^+$ and O$_2^+$. 
Notably, with TI coherence, the $B$ state of N$_2^+$ exhibits a higher population than the $X$ state in a wide range of angles. 
This suggests that TI coherence may play a crucial role in explaining population inversion phenomena observed in N$_2^+$ lasing. 
Our predictions for N$_2^+$ could be tested in ATAS experiments, while simulations and experiments of strong-field dissociation of O$_2^+$ can verify the results for O$_2^+$.

Regarding SDI, we found that TI coherence does not alter the qualitative behavior of the KER spectra, and the quantitative differences between results with or without TI coherence are too small to be resolved experimentally. 
However, the angle-dependent yields of the dication show clear signatures of TI coherence. 
Therefore, when studying angular-averaged quantities such as KER spectra, the model can be simplified, as the detailed dynamics are likely to be washed out.

In the future, we expect that TI coherence will exert a more pronounced impact on experiments involving few-cycle intense IR pulses for single ionization compared to SDI.
This is due to the difficulty in controlling the shape of few-cycle IR pulses, which may contain pre-pulses and post-pulses with intensities on the order of 10--20\% of the main peak intensity. 
In SDI, the peak intensity is typically around $10^{15}$ W/cm$^2$,  such that pre-pulses may have intensities of $1 - 2 \times 10^{14}$ W/cm$^2$,  leading to single ionization of neutral molecules and resulting in small quantitative changes in the final observables. 
In addition,  the experimental uncertainty in determining the peak intensity of such intense laser pulses is generally higher than that of weak pulses. 
Considering these uncertainties, the influence of TI coherence is expected to be weaker in SDI compared to single ionization. 
Therefore,  we recommend that future models of multiorbital single ionization of molecules should incorporate TI coherence in the equations of motion, while for SDI, it is reasonable to neglect TI coherence to further simplify the model.

\begin{acknowledgments}
This work was supported by Chemical Sciences, Geosciences and Biosciences Division, Office of Basic Energy Sciences, Office of Science, U.S. Department of Energy under Grant No. DE-FG02-86ER13491.
\end{acknowledgments}

%


\begin{thebibliography}{34}%
\makeatletter
\providecommand \@ifxundefined [1]{%
 \@ifx{#1\undefined}
}%
\providecommand \@ifnum [1]{%
 \ifnum #1\expandafter \@firstoftwo
 \else \expandafter \@secondoftwo
 \fi
}%
\providecommand \@ifx [1]{%
 \ifx #1\expandafter \@firstoftwo
 \else \expandafter \@secondoftwo
 \fi
}%
\providecommand \natexlab [1]{#1}%
\providecommand \enquote  [1]{``#1''}%
\providecommand \bibnamefont  [1]{#1}%
\providecommand \bibfnamefont [1]{#1}%
\providecommand \citenamefont [1]{#1}%
\providecommand \href@noop [0]{\@secondoftwo}%
\providecommand \href [0]{\begingroup \@sanitize@url \@href}%
\providecommand \@href[1]{\@@startlink{#1}\@@href}%
\providecommand \@@href[1]{\endgroup#1\@@endlink}%
\providecommand \@sanitize@url [0]{\catcode `\\12\catcode `\$12\catcode
  `\&12\catcode `\#12\catcode `\^12\catcode `\_12\catcode `\%12\relax}%
\providecommand \@@startlink[1]{}%
\providecommand \@@endlink[0]{}%
\providecommand \url  [0]{\begingroup\@sanitize@url \@url }%
\providecommand \@url [1]{\endgroup\@href {#1}{\urlprefix }}%
\providecommand \urlprefix  [0]{URL }%
\providecommand \Eprint [0]{\href }%
\providecommand \doibase [0]{https://doi.org/}%
\providecommand \selectlanguage [0]{\@gobble}%
\providecommand \bibinfo  [0]{\@secondoftwo}%
\providecommand \bibfield  [0]{\@secondoftwo}%
\providecommand \translation [1]{[#1]}%
\providecommand \BibitemOpen [0]{}%
\providecommand \bibitemStop [0]{}%
\providecommand \bibitemNoStop [0]{.\EOS\space}%
\providecommand \EOS [0]{\spacefactor3000\relax}%
\providecommand \BibitemShut  [1]{\csname bibitem#1\endcsname}%
\let\auto@bib@innerbib\@empty
\bibitem [{\citenamefont {Voss}\ \emph {et~al.}(2004)\citenamefont {Voss},
  \citenamefont {Alnaser}, \citenamefont {Tong}, \citenamefont {Maharjan},
  \citenamefont {Ranitovic}, \citenamefont {Ulrich}, \citenamefont {Shan},
  \citenamefont {Chang}, \citenamefont {Lin},\ and\ \citenamefont
  {Cocke}}]{voss2004}%
  \BibitemOpen
  \bibfield  {author} {\bibinfo {author} {\bibfnamefont {S.}~\bibnamefont
  {Voss}}, \bibinfo {author} {\bibfnamefont {A.~S.}\ \bibnamefont {Alnaser}},
  \bibinfo {author} {\bibfnamefont {X.~M.}\ \bibnamefont {Tong}}, \bibinfo
  {author} {\bibfnamefont {C.}~\bibnamefont {Maharjan}}, \bibinfo {author}
  {\bibfnamefont {P.}~\bibnamefont {Ranitovic}}, \bibinfo {author}
  {\bibfnamefont {B.}~\bibnamefont {Ulrich}}, \bibinfo {author} {\bibfnamefont
  {B.}~\bibnamefont {Shan}}, \bibinfo {author} {\bibfnamefont {Z.}~\bibnamefont
  {Chang}}, \bibinfo {author} {\bibfnamefont {C.~D.}\ \bibnamefont {Lin}},\
  and\ \bibinfo {author} {\bibfnamefont {C.~L.}\ \bibnamefont {Cocke}},\ }\href
  {https://doi.org/https://doi.org/10.1088/0953-4075/37/21/002} {\bibfield
  {journal} {\bibinfo  {journal} {J. Phys. B: At. Mol. Opt. Phys.}\ }\textbf
  {\bibinfo {volume} {37}},\ \bibinfo {pages} {4239} (\bibinfo {year}
  {2004})}\BibitemShut {NoStop}%
\bibitem [{\citenamefont {Wu}\ \emph {et~al.}(2011)\citenamefont {Wu},
  \citenamefont {Yang}, \citenamefont {Wu}, \citenamefont {Chen}, \citenamefont
  {Dong}, \citenamefont {Liu}, \citenamefont {Deng}, \citenamefont {Liu},
  \citenamefont {Liu},\ and\ \citenamefont {Gong}}]{wu2011}%
  \BibitemOpen
  \bibfield  {author} {\bibinfo {author} {\bibfnamefont {C.}~\bibnamefont
  {Wu}}, \bibinfo {author} {\bibfnamefont {Y.}~\bibnamefont {Yang}}, \bibinfo
  {author} {\bibfnamefont {Z.}~\bibnamefont {Wu}}, \bibinfo {author}
  {\bibfnamefont {B.}~\bibnamefont {Chen}}, \bibinfo {author} {\bibfnamefont
  {H.}~\bibnamefont {Dong}}, \bibinfo {author} {\bibfnamefont {X.}~\bibnamefont
  {Liu}}, \bibinfo {author} {\bibfnamefont {Y.}~\bibnamefont {Deng}}, \bibinfo
  {author} {\bibfnamefont {H.}~\bibnamefont {Liu}}, \bibinfo {author}
  {\bibfnamefont {Y.}~\bibnamefont {Liu}},\ and\ \bibinfo {author}
  {\bibfnamefont {Q.}~\bibnamefont {Gong}},\ }\href
  {https://doi.org/https://doi.org/10.1039/C1CP21345H} {\bibfield  {journal}
  {\bibinfo  {journal} {Phys. Chem. Chem. Phys.}\ }\textbf {\bibinfo {volume}
  {13}},\ \bibinfo {pages} {18398} (\bibinfo {year} {2011})}\BibitemShut
  {NoStop}%
\bibitem [{\citenamefont {Wu}\ \emph {et~al.}(2012)\citenamefont {Wu},
  \citenamefont {Schmidt}, \citenamefont {Kunitski}, \citenamefont {Meckel},
  \citenamefont {Voss}, \citenamefont {Sann}, \citenamefont {Kim},
  \citenamefont {Jahnke}, \citenamefont {Czasch},\ and\ \citenamefont
  {D{\"o}rner}}]{wu2012}%
  \BibitemOpen
  \bibfield  {author} {\bibinfo {author} {\bibfnamefont {J.}~\bibnamefont
  {Wu}}, \bibinfo {author} {\bibfnamefont {L.~P.~H.}\ \bibnamefont {Schmidt}},
  \bibinfo {author} {\bibfnamefont {M.}~\bibnamefont {Kunitski}}, \bibinfo
  {author} {\bibfnamefont {M.}~\bibnamefont {Meckel}}, \bibinfo {author}
  {\bibfnamefont {S.}~\bibnamefont {Voss}}, \bibinfo {author} {\bibfnamefont
  {H.}~\bibnamefont {Sann}}, \bibinfo {author} {\bibfnamefont {H.}~\bibnamefont
  {Kim}}, \bibinfo {author} {\bibfnamefont {T.}~\bibnamefont {Jahnke}},
  \bibinfo {author} {\bibfnamefont {A.}~\bibnamefont {Czasch}},\ and\ \bibinfo
  {author} {\bibfnamefont {R.}~\bibnamefont {D{\"o}rner}},\ }\href
  {https://doi.org/https://doi.org/10.1103/PhysRevLett.108.183001} {\bibfield
  {journal} {\bibinfo  {journal} {Phys. Rev. Lett.}\ }\textbf {\bibinfo
  {volume} {108}},\ \bibinfo {pages} {183001} (\bibinfo {year}
  {2012})}\BibitemShut {NoStop}%
\bibitem [{\citenamefont {McFarland}\ \emph {et~al.}(2008)\citenamefont
  {McFarland}, \citenamefont {Farrell}, \citenamefont {Bucksbaum},\ and\
  \citenamefont {Guhr}}]{mcfarland2008}%
  \BibitemOpen
  \bibfield  {author} {\bibinfo {author} {\bibfnamefont {B.~K.}\ \bibnamefont
  {McFarland}}, \bibinfo {author} {\bibfnamefont {J.~P.}\ \bibnamefont
  {Farrell}}, \bibinfo {author} {\bibfnamefont {P.~H.}\ \bibnamefont
  {Bucksbaum}},\ and\ \bibinfo {author} {\bibfnamefont {M.}~\bibnamefont
  {Guhr}},\ }\href {https://doi.org/https://doi.org/10.1126/science.1162780}
  {\bibfield  {journal} {\bibinfo  {journal} {Science}\ }\textbf {\bibinfo
  {volume} {322}},\ \bibinfo {pages} {1232} (\bibinfo {year}
  {2008})}\BibitemShut {NoStop}%
\bibitem [{\citenamefont {Smirnova}\ \emph {et~al.}(2009)\citenamefont
  {Smirnova}, \citenamefont {Mairesse}, \citenamefont {Patchkovskii},
  \citenamefont {Dudovich}, \citenamefont {Villeneuve}, \citenamefont
  {Corkum},\ and\ \citenamefont {Ivanov}}]{smirnova2009}%
  \BibitemOpen
  \bibfield  {author} {\bibinfo {author} {\bibfnamefont {O.}~\bibnamefont
  {Smirnova}}, \bibinfo {author} {\bibfnamefont {Y.}~\bibnamefont {Mairesse}},
  \bibinfo {author} {\bibfnamefont {S.}~\bibnamefont {Patchkovskii}}, \bibinfo
  {author} {\bibfnamefont {N.}~\bibnamefont {Dudovich}}, \bibinfo {author}
  {\bibfnamefont {D.}~\bibnamefont {Villeneuve}}, \bibinfo {author}
  {\bibfnamefont {P.}~\bibnamefont {Corkum}},\ and\ \bibinfo {author}
  {\bibfnamefont {M.~Y.}\ \bibnamefont {Ivanov}},\ }\href
  {https://doi.org/https://doi.org/10.1038/nature08253} {\bibfield  {journal}
  {\bibinfo  {journal} {Nature}\ }\textbf {\bibinfo {volume} {460}},\ \bibinfo
  {pages} {972} (\bibinfo {year} {2009})}\BibitemShut {NoStop}%
\bibitem [{\citenamefont {Farrell}\ \emph {et~al.}(2011)\citenamefont
  {Farrell}, \citenamefont {Petretti}, \citenamefont {F{\"o}rster},
  \citenamefont {McFarland}, \citenamefont {Spector}, \citenamefont {Vanne},
  \citenamefont {Decleva}, \citenamefont {Bucksbaum}, \citenamefont {Saenz},\
  and\ \citenamefont {G{\"u}hr}}]{farrell2011}%
  \BibitemOpen
  \bibfield  {author} {\bibinfo {author} {\bibfnamefont {J.~P.}\ \bibnamefont
  {Farrell}}, \bibinfo {author} {\bibfnamefont {S.}~\bibnamefont {Petretti}},
  \bibinfo {author} {\bibfnamefont {J.}~\bibnamefont {F{\"o}rster}}, \bibinfo
  {author} {\bibfnamefont {B.~K.}\ \bibnamefont {McFarland}}, \bibinfo {author}
  {\bibfnamefont {L.~S.}\ \bibnamefont {Spector}}, \bibinfo {author}
  {\bibfnamefont {Y.~V.}\ \bibnamefont {Vanne}}, \bibinfo {author}
  {\bibfnamefont {P.}~\bibnamefont {Decleva}}, \bibinfo {author} {\bibfnamefont
  {P.~H.}\ \bibnamefont {Bucksbaum}}, \bibinfo {author} {\bibfnamefont
  {A.}~\bibnamefont {Saenz}},\ and\ \bibinfo {author} {\bibfnamefont
  {M.}~\bibnamefont {G{\"u}hr}},\ }\href
  {https://doi.org/https://doi.org/10.1103/PhysRevLett.107.083001} {\bibfield
  {journal} {\bibinfo  {journal} {Phys. Rev. Lett.}\ }\textbf {\bibinfo
  {volume} {107}},\ \bibinfo {pages} {083001} (\bibinfo {year}
  {2011})}\BibitemShut {NoStop}%
\bibitem [{\citenamefont {He}\ \emph {et~al.}(2022)\citenamefont {He},
  \citenamefont {Sun}, \citenamefont {Lan}, \citenamefont {He}, \citenamefont
  {Wang}, \citenamefont {Wang}, \citenamefont {Zhu}, \citenamefont {Li},
  \citenamefont {Cao}, \citenamefont {Lu},\ and\ \citenamefont {Lin}}]{He2022}%
  \BibitemOpen
  \bibfield  {author} {\bibinfo {author} {\bibfnamefont {L.}~\bibnamefont
  {He}}, \bibinfo {author} {\bibfnamefont {S.}~\bibnamefont {Sun}}, \bibinfo
  {author} {\bibfnamefont {P.}~\bibnamefont {Lan}}, \bibinfo {author}
  {\bibfnamefont {Y.}~\bibnamefont {He}}, \bibinfo {author} {\bibfnamefont
  {B.}~\bibnamefont {Wang}}, \bibinfo {author} {\bibfnamefont {P.}~\bibnamefont
  {Wang}}, \bibinfo {author} {\bibfnamefont {X.}~\bibnamefont {Zhu}}, \bibinfo
  {author} {\bibfnamefont {L.}~\bibnamefont {Li}}, \bibinfo {author}
  {\bibfnamefont {W.}~\bibnamefont {Cao}}, \bibinfo {author} {\bibfnamefont
  {P.}~\bibnamefont {Lu}},\ and\ \bibinfo {author} {\bibfnamefont {C.~D.}\
  \bibnamefont {Lin}},\ }\href
  {https://doi.org/https://doi.org/10.1038/s41467-022-32313-0} {\bibfield
  {journal} {\bibinfo  {journal} {Nat. commun.}\ }\textbf {\bibinfo {volume}
  {13}},\ \bibinfo {pages} {1} (\bibinfo {year} {2022})}\BibitemShut {NoStop}%
\bibitem [{\citenamefont {Akagi}\ \emph {et~al.}(2009)\citenamefont {Akagi},
  \citenamefont {Otobe}, \citenamefont {Staudte}, \citenamefont {Shiner},
  \citenamefont {Turner}, \citenamefont {D{\"o}rner}, \citenamefont
  {Villeneuve},\ and\ \citenamefont {Corkum}}]{akagi2009}%
  \BibitemOpen
  \bibfield  {author} {\bibinfo {author} {\bibfnamefont {H.}~\bibnamefont
  {Akagi}}, \bibinfo {author} {\bibfnamefont {T.}~\bibnamefont {Otobe}},
  \bibinfo {author} {\bibfnamefont {A.}~\bibnamefont {Staudte}}, \bibinfo
  {author} {\bibfnamefont {A.}~\bibnamefont {Shiner}}, \bibinfo {author}
  {\bibfnamefont {F.}~\bibnamefont {Turner}}, \bibinfo {author} {\bibfnamefont
  {R.}~\bibnamefont {D{\"o}rner}}, \bibinfo {author} {\bibfnamefont
  {D.}~\bibnamefont {Villeneuve}},\ and\ \bibinfo {author} {\bibfnamefont
  {P.}~\bibnamefont {Corkum}},\ }\href
  {https://doi.org/https://doi.org/10.1126/science.1175253} {\bibfield
  {journal} {\bibinfo  {journal} {Science}\ }\textbf {\bibinfo {volume}
  {325}},\ \bibinfo {pages} {1364} (\bibinfo {year} {2009})}\BibitemShut
  {NoStop}%
\bibitem [{\citenamefont {De}\ \emph {et~al.}(2011)\citenamefont {De},
  \citenamefont {Magrakvelidze}, \citenamefont {Bocharova}, \citenamefont
  {Ray}, \citenamefont {Cao}, \citenamefont {Znakovskaya}, \citenamefont {Li},
  \citenamefont {Wang}, \citenamefont {Laurent}, \citenamefont {Thumm},
  \citenamefont {Kling}, \citenamefont {Litvinyuk}, \citenamefont
  {Ben-Itzhak},\ and\ \citenamefont {Cocke}}]{de2011}%
  \BibitemOpen
  \bibfield  {author} {\bibinfo {author} {\bibfnamefont {S.}~\bibnamefont
  {De}}, \bibinfo {author} {\bibfnamefont {M.}~\bibnamefont {Magrakvelidze}},
  \bibinfo {author} {\bibfnamefont {I.~A.}\ \bibnamefont {Bocharova}}, \bibinfo
  {author} {\bibfnamefont {D.}~\bibnamefont {Ray}}, \bibinfo {author}
  {\bibfnamefont {W.}~\bibnamefont {Cao}}, \bibinfo {author} {\bibfnamefont
  {I.}~\bibnamefont {Znakovskaya}}, \bibinfo {author} {\bibfnamefont
  {H.}~\bibnamefont {Li}}, \bibinfo {author} {\bibfnamefont {Z.}~\bibnamefont
  {Wang}}, \bibinfo {author} {\bibfnamefont {G.}~\bibnamefont {Laurent}},
  \bibinfo {author} {\bibfnamefont {U.}~\bibnamefont {Thumm}}, \bibinfo
  {author} {\bibfnamefont {M.~F.}\ \bibnamefont {Kling}}, \bibinfo {author}
  {\bibfnamefont {I.~V.}\ \bibnamefont {Litvinyuk}}, \bibinfo {author}
  {\bibfnamefont {I.}~\bibnamefont {Ben-Itzhak}},\ and\ \bibinfo {author}
  {\bibfnamefont {C.~L.}\ \bibnamefont {Cocke}},\ }\href
  {https://doi.org/https://doi.org/10.1103/PhysRevA.84.043410} {\bibfield
  {journal} {\bibinfo  {journal} {Phys. Rev. A}\ }\textbf {\bibinfo {volume}
  {84}},\ \bibinfo {pages} {043410} (\bibinfo {year} {2011})}\BibitemShut
  {NoStop}%
\bibitem [{\citenamefont {Liu}\ \emph {et~al.}(2013)\citenamefont {Liu},
  \citenamefont {Brelet}, \citenamefont {Point}, \citenamefont {Houard},\ and\
  \citenamefont {Mysyrowicz}}]{liu2013}%
  \BibitemOpen
  \bibfield  {author} {\bibinfo {author} {\bibfnamefont {Y.}~\bibnamefont
  {Liu}}, \bibinfo {author} {\bibfnamefont {Y.}~\bibnamefont {Brelet}},
  \bibinfo {author} {\bibfnamefont {G.}~\bibnamefont {Point}}, \bibinfo
  {author} {\bibfnamefont {A.}~\bibnamefont {Houard}},\ and\ \bibinfo {author}
  {\bibfnamefont {A.}~\bibnamefont {Mysyrowicz}},\ }\href
  {https://doi.org/https://doi.org/10.1364/OE.21.022791} {\bibfield  {journal}
  {\bibinfo  {journal} {Opt. Express}\ }\textbf {\bibinfo {volume} {21}},\
  \bibinfo {pages} {22791} (\bibinfo {year} {2013})}\BibitemShut {NoStop}%
\bibitem [{\citenamefont {Xu}\ \emph {et~al.}(2015)\citenamefont {Xu},
  \citenamefont {L{\"o}tstedt}, \citenamefont {Iwasaki},\ and\ \citenamefont
  {Yamanouchi}}]{xu2015}%
  \BibitemOpen
  \bibfield  {author} {\bibinfo {author} {\bibfnamefont {H.}~\bibnamefont
  {Xu}}, \bibinfo {author} {\bibfnamefont {E.}~\bibnamefont {L{\"o}tstedt}},
  \bibinfo {author} {\bibfnamefont {A.}~\bibnamefont {Iwasaki}},\ and\ \bibinfo
  {author} {\bibfnamefont {K.}~\bibnamefont {Yamanouchi}},\ }\href
  {https://doi.org/https://doi.org/10.1038/ncomms9347} {\bibfield  {journal}
  {\bibinfo  {journal} {Nat. Commun.}\ }\textbf {\bibinfo {volume} {6}},\
  \bibinfo {pages} {8347} (\bibinfo {year} {2015})}\BibitemShut {NoStop}%
\bibitem [{\citenamefont {Yao}\ \emph {et~al.}(2016)\citenamefont {Yao},
  \citenamefont {Jiang}, \citenamefont {Chu}, \citenamefont {Zeng},
  \citenamefont {Wu}, \citenamefont {Lu}, \citenamefont {Li}, \citenamefont
  {Xie}, \citenamefont {Li}, \citenamefont {Yu} \emph {et~al.}}]{yao2016}%
  \BibitemOpen
  \bibfield  {author} {\bibinfo {author} {\bibfnamefont {J.}~\bibnamefont
  {Yao}}, \bibinfo {author} {\bibfnamefont {S.}~\bibnamefont {Jiang}}, \bibinfo
  {author} {\bibfnamefont {W.}~\bibnamefont {Chu}}, \bibinfo {author}
  {\bibfnamefont {B.}~\bibnamefont {Zeng}}, \bibinfo {author} {\bibfnamefont
  {C.}~\bibnamefont {Wu}}, \bibinfo {author} {\bibfnamefont {R.}~\bibnamefont
  {Lu}}, \bibinfo {author} {\bibfnamefont {Z.}~\bibnamefont {Li}}, \bibinfo
  {author} {\bibfnamefont {H.}~\bibnamefont {Xie}}, \bibinfo {author}
  {\bibfnamefont {G.}~\bibnamefont {Li}}, \bibinfo {author} {\bibfnamefont
  {C.}~\bibnamefont {Yu}}, \emph {et~al.},\ }\href
  {https://doi.org/https://doi.org/10.1103/PhysRevLett.116.143007} {\bibfield
  {journal} {\bibinfo  {journal} {Phys. Rev. Lett.}\ }\textbf {\bibinfo
  {volume} {116}},\ \bibinfo {pages} {143007} (\bibinfo {year}
  {2016})}\BibitemShut {NoStop}%
\bibitem [{\citenamefont {Kleine}\ \emph {et~al.}(2022)\citenamefont {Kleine},
  \citenamefont {Winghart}, \citenamefont {Zhang}, \citenamefont {Richter},
  \citenamefont {Ekimova}, \citenamefont {Eckert}, \citenamefont {Vrakking},
  \citenamefont {Nibbering}, \citenamefont {Rouz{\'e}e},\ and\ \citenamefont
  {Grant}}]{kleine2022}%
  \BibitemOpen
  \bibfield  {author} {\bibinfo {author} {\bibfnamefont {C.}~\bibnamefont
  {Kleine}}, \bibinfo {author} {\bibfnamefont {M.-O.}\ \bibnamefont
  {Winghart}}, \bibinfo {author} {\bibfnamefont {Z.-Y.}\ \bibnamefont {Zhang}},
  \bibinfo {author} {\bibfnamefont {M.}~\bibnamefont {Richter}}, \bibinfo
  {author} {\bibfnamefont {M.}~\bibnamefont {Ekimova}}, \bibinfo {author}
  {\bibfnamefont {S.}~\bibnamefont {Eckert}}, \bibinfo {author} {\bibfnamefont
  {M.~J.}\ \bibnamefont {Vrakking}}, \bibinfo {author} {\bibfnamefont {E.~T.}\
  \bibnamefont {Nibbering}}, \bibinfo {author} {\bibfnamefont {A.}~\bibnamefont
  {Rouz{\'e}e}},\ and\ \bibinfo {author} {\bibfnamefont {E.~R.}\ \bibnamefont
  {Grant}},\ }\href
  {https://doi.org/https://doi.org/10.1103/PhysRevLett.129.123002} {\bibfield
  {journal} {\bibinfo  {journal} {Phys. Rev. Lett.}\ }\textbf {\bibinfo
  {volume} {129}},\ \bibinfo {pages} {123002} (\bibinfo {year}
  {2022})}\BibitemShut {NoStop}%
\bibitem [{\citenamefont {Zhang}\ \emph {et~al.}(2020)\citenamefont {Zhang},
  \citenamefont {Xie}, \citenamefont {Li}, \citenamefont {Wang}, \citenamefont
  {Lei}, \citenamefont {Zhao}, \citenamefont {Chen}, \citenamefont {Yao},
  \citenamefont {Cheng},\ and\ \citenamefont {Zhao}}]{zhang2020}%
  \BibitemOpen
  \bibfield  {author} {\bibinfo {author} {\bibfnamefont {Q.}~\bibnamefont
  {Zhang}}, \bibinfo {author} {\bibfnamefont {H.}~\bibnamefont {Xie}}, \bibinfo
  {author} {\bibfnamefont {G.}~\bibnamefont {Li}}, \bibinfo {author}
  {\bibfnamefont {X.}~\bibnamefont {Wang}}, \bibinfo {author} {\bibfnamefont
  {H.}~\bibnamefont {Lei}}, \bibinfo {author} {\bibfnamefont {J.}~\bibnamefont
  {Zhao}}, \bibinfo {author} {\bibfnamefont {Z.}~\bibnamefont {Chen}}, \bibinfo
  {author} {\bibfnamefont {J.}~\bibnamefont {Yao}}, \bibinfo {author}
  {\bibfnamefont {Y.}~\bibnamefont {Cheng}},\ and\ \bibinfo {author}
  {\bibfnamefont {Z.}~\bibnamefont {Zhao}},\ }\href
  {https://doi.org/https://doi.org/10.1038/s42005-020-0321-7} {\bibfield
  {journal} {\bibinfo  {journal} {Commun. Phys.}\ }\textbf {\bibinfo {volume}
  {3}},\ \bibinfo {pages} {1} (\bibinfo {year} {2020})}\BibitemShut {NoStop}%
\bibitem [{\citenamefont {Yuen}\ and\ \citenamefont {Lin}(2022)}]{yuen2022}%
  \BibitemOpen
  \bibfield  {author} {\bibinfo {author} {\bibfnamefont {C.~H.}\ \bibnamefont
  {Yuen}}\ and\ \bibinfo {author} {\bibfnamefont {C.~D.}\ \bibnamefont {Lin}},\
  }\href {https://doi.org/https://doi.org/10.1103/PhysRevA.106.023120}
  {\bibfield  {journal} {\bibinfo  {journal} {Phys. Rev. A}\ }\textbf {\bibinfo
  {volume} {106}},\ \bibinfo {pages} {023120} (\bibinfo {year}
  {2022})}\BibitemShut {NoStop}%
\bibitem [{\citenamefont {Yuen}\ \emph {et~al.}(2023)\citenamefont {Yuen},
  \citenamefont {Modak}, \citenamefont {Song}, \citenamefont {Zhao},\ and\
  \citenamefont {Lin}}]{yuen2023}%
  \BibitemOpen
  \bibfield  {author} {\bibinfo {author} {\bibfnamefont {C.~H.}\ \bibnamefont
  {Yuen}}, \bibinfo {author} {\bibfnamefont {P.}~\bibnamefont {Modak}},
  \bibinfo {author} {\bibfnamefont {Y.}~\bibnamefont {Song}}, \bibinfo {author}
  {\bibfnamefont {S.-F.}\ \bibnamefont {Zhao}},\ and\ \bibinfo {author}
  {\bibfnamefont {C.~D.}\ \bibnamefont {Lin}},\ }\href
  {https://doi.org/https://doi.org/10.1103/PhysRevA.107.013112} {\bibfield
  {journal} {\bibinfo  {journal} {Phys. Rev. A}\ }\textbf {\bibinfo {volume}
  {107}},\ \bibinfo {pages} {013112} (\bibinfo {year} {2023})}\BibitemShut
  {NoStop}%
\bibitem [{\citenamefont {Rohringer}\ and\ \citenamefont
  {Santra}(2009)}]{rohringer2009}%
  \BibitemOpen
  \bibfield  {author} {\bibinfo {author} {\bibfnamefont {N.}~\bibnamefont
  {Rohringer}}\ and\ \bibinfo {author} {\bibfnamefont {R.}~\bibnamefont
  {Santra}},\ }\href
  {https://doi.org/https://doi.org/10.1103/PhysRevA.79.053402} {\bibfield
  {journal} {\bibinfo  {journal} {Phys. Rev. A}\ }\textbf {\bibinfo {volume}
  {79}},\ \bibinfo {pages} {053402} (\bibinfo {year} {2009})}\BibitemShut
  {NoStop}%
\bibitem [{\citenamefont {Goulielmakis}\ \emph {et~al.}(2010)\citenamefont
  {Goulielmakis}, \citenamefont {Loh}, \citenamefont {Wirth}, \citenamefont
  {Santra}, \citenamefont {Rohringer}, \citenamefont {Yakovlev}, \citenamefont
  {Zherebtsov}, \citenamefont {Pfeifer}, \citenamefont {Azzeer}, \citenamefont
  {Kling}, \citenamefont {Leone},\ and\ \citenamefont
  {Krausz}}]{goulielmakis2010}%
  \BibitemOpen
  \bibfield  {author} {\bibinfo {author} {\bibfnamefont {E.}~\bibnamefont
  {Goulielmakis}}, \bibinfo {author} {\bibfnamefont {Z.-H.}\ \bibnamefont
  {Loh}}, \bibinfo {author} {\bibfnamefont {A.}~\bibnamefont {Wirth}}, \bibinfo
  {author} {\bibfnamefont {R.}~\bibnamefont {Santra}}, \bibinfo {author}
  {\bibfnamefont {N.}~\bibnamefont {Rohringer}}, \bibinfo {author}
  {\bibfnamefont {V.~S.}\ \bibnamefont {Yakovlev}}, \bibinfo {author}
  {\bibfnamefont {S.}~\bibnamefont {Zherebtsov}}, \bibinfo {author}
  {\bibfnamefont {T.}~\bibnamefont {Pfeifer}}, \bibinfo {author} {\bibfnamefont
  {A.~M.}\ \bibnamefont {Azzeer}}, \bibinfo {author} {\bibfnamefont {M.~F.}\
  \bibnamefont {Kling}}, \bibinfo {author} {\bibfnamefont {S.}~\bibnamefont
  {Leone}},\ and\ \bibinfo {author} {\bibfnamefont {F.}~\bibnamefont
  {Krausz}},\ }\href {https://doi.org/https://doi.org/10.1038/nature09212}
  {\bibfield  {journal} {\bibinfo  {journal} {Nature}\ }\textbf {\bibinfo
  {volume} {466}},\ \bibinfo {pages} {739} (\bibinfo {year}
  {2010})}\BibitemShut {NoStop}%
\bibitem [{\citenamefont {Pabst}\ \emph {et~al.}(2016)\citenamefont {Pabst},
  \citenamefont {Lein},\ and\ \citenamefont {W{\"o}rner}}]{pabst2016}%
  \BibitemOpen
  \bibfield  {author} {\bibinfo {author} {\bibfnamefont {S.}~\bibnamefont
  {Pabst}}, \bibinfo {author} {\bibfnamefont {M.}~\bibnamefont {Lein}},\ and\
  \bibinfo {author} {\bibfnamefont {H.~J.}\ \bibnamefont {W{\"o}rner}},\ }\href
  {https://doi.org/https://doi.org/10.1103/PhysRevA.93.023412} {\bibfield
  {journal} {\bibinfo  {journal} {Phys. Rev. A}\ }\textbf {\bibinfo {volume}
  {93}},\ \bibinfo {pages} {023412} (\bibinfo {year} {2016})}\BibitemShut
  {NoStop}%
\bibitem [{\citenamefont {Xue}\ \emph {et~al.}(2021)\citenamefont {Xue},
  \citenamefont {Yue}, \citenamefont {Du}, \citenamefont {Hu},\ and\
  \citenamefont {Le}}]{xue2021}%
  \BibitemOpen
  \bibfield  {author} {\bibinfo {author} {\bibfnamefont {S.}~\bibnamefont
  {Xue}}, \bibinfo {author} {\bibfnamefont {S.}~\bibnamefont {Yue}}, \bibinfo
  {author} {\bibfnamefont {H.}~\bibnamefont {Du}}, \bibinfo {author}
  {\bibfnamefont {B.}~\bibnamefont {Hu}},\ and\ \bibinfo {author}
  {\bibfnamefont {A.-T.}\ \bibnamefont {Le}},\ }\href
  {https://doi.org/https://doi.org/10.1103/PhysRevA.104.013101} {\bibfield
  {journal} {\bibinfo  {journal} {Phys. Rev. A}\ }\textbf {\bibinfo {volume}
  {104}},\ \bibinfo {pages} {013101} (\bibinfo {year} {2021})}\BibitemShut
  {NoStop}%
\bibitem [{\citenamefont {Xue}\ \emph {et~al.}(2022)\citenamefont {Xue},
  \citenamefont {Sun}, \citenamefont {Ding}, \citenamefont {Hu}, \citenamefont
  {Yue},\ and\ \citenamefont {Du}}]{xue2022}%
  \BibitemOpen
  \bibfield  {author} {\bibinfo {author} {\bibfnamefont {S.}~\bibnamefont
  {Xue}}, \bibinfo {author} {\bibfnamefont {S.}~\bibnamefont {Sun}}, \bibinfo
  {author} {\bibfnamefont {P.}~\bibnamefont {Ding}}, \bibinfo {author}
  {\bibfnamefont {B.}~\bibnamefont {Hu}}, \bibinfo {author} {\bibfnamefont
  {S.}~\bibnamefont {Yue}},\ and\ \bibinfo {author} {\bibfnamefont
  {H.}~\bibnamefont {Du}},\ }\href
  {https://doi.org/https://doi.org/10.1103/PhysRevA.105.043108} {\bibfield
  {journal} {\bibinfo  {journal} {Phys. Rev. A}\ }\textbf {\bibinfo {volume}
  {105}},\ \bibinfo {pages} {043108} (\bibinfo {year} {2022})}\BibitemShut
  {NoStop}%
\bibitem [{\citenamefont {Cederbaum}\ and\ \citenamefont
  {Zobeley}(1999)}]{cederbaum1999}%
  \BibitemOpen
  \bibfield  {author} {\bibinfo {author} {\bibfnamefont {L.~S.}\ \bibnamefont
  {Cederbaum}}\ and\ \bibinfo {author} {\bibfnamefont {J.}~\bibnamefont
  {Zobeley}},\ }\href
  {https://doi.org/https://doi.org/10.1016/S0009-2614(99)00508-4} {\bibfield
  {journal} {\bibinfo  {journal} {Chem. Phys. Lett.}\ }\textbf {\bibinfo
  {volume} {307}},\ \bibinfo {pages} {205} (\bibinfo {year}
  {1999})}\BibitemShut {NoStop}%
\bibitem [{\citenamefont {L{\'e}pine}\ \emph {et~al.}(2014)\citenamefont
  {L{\'e}pine}, \citenamefont {Ivanov},\ and\ \citenamefont
  {Vrakking}}]{lepine2014}%
  \BibitemOpen
  \bibfield  {author} {\bibinfo {author} {\bibfnamefont {F.}~\bibnamefont
  {L{\'e}pine}}, \bibinfo {author} {\bibfnamefont {M.~Y.}\ \bibnamefont
  {Ivanov}},\ and\ \bibinfo {author} {\bibfnamefont {M.~J.}\ \bibnamefont
  {Vrakking}},\ }\href
  {https://doi.org/https://doi.org/10.1038/nphoton.2014.25} {\bibfield
  {journal} {\bibinfo  {journal} {Nat. Photonics}\ }\textbf {\bibinfo {volume}
  {8}},\ \bibinfo {pages} {195} (\bibinfo {year} {2014})}\BibitemShut {NoStop}%
\bibitem [{\citenamefont {Yuen}\ and\ \citenamefont {Lin}(2023)}]{yuen2023b}%
  \BibitemOpen
  \bibfield  {author} {\bibinfo {author} {\bibfnamefont {C.~H.}\ \bibnamefont
  {Yuen}}\ and\ \bibinfo {author} {\bibfnamefont {C.~D.}\ \bibnamefont {Lin}},\
  }\bibfield  {journal} {\bibinfo  {journal} {arXiv:2302.05944}\ }\href
  {https://doi.org/https://doi.org/10.48550/arXiv.2302.05944}
  {https://doi.org/10.48550/arXiv.2302.05944} (\bibinfo {year}
  {2023})\BibitemShut {NoStop}%
\bibitem [{\citenamefont {Arnold}\ \emph {et~al.}(2020)\citenamefont {Arnold},
  \citenamefont {Larivi{\`e}re-Loiselle}, \citenamefont {Khalili},
  \citenamefont {Inhester}, \citenamefont {Welsch},\ and\ \citenamefont
  {Santra}}]{arnold2020}%
  \BibitemOpen
  \bibfield  {author} {\bibinfo {author} {\bibfnamefont {C.}~\bibnamefont
  {Arnold}}, \bibinfo {author} {\bibfnamefont {C.}~\bibnamefont
  {Larivi{\`e}re-Loiselle}}, \bibinfo {author} {\bibfnamefont {K.}~\bibnamefont
  {Khalili}}, \bibinfo {author} {\bibfnamefont {L.}~\bibnamefont {Inhester}},
  \bibinfo {author} {\bibfnamefont {R.}~\bibnamefont {Welsch}},\ and\ \bibinfo
  {author} {\bibfnamefont {R.}~\bibnamefont {Santra}},\ }\href
  {https://doi.org/https://doi.org/10.1088/1361-6455/ab9658} {\bibfield
  {journal} {\bibinfo  {journal} {J. Phys. B: At., Mol. Opt. Phys.}\ }\textbf
  {\bibinfo {volume} {53}},\ \bibinfo {pages} {164006} (\bibinfo {year}
  {2020})}\BibitemShut {NoStop}%
\bibitem [{\citenamefont {Tolstikhin}\ \emph {et~al.}(2011)\citenamefont
  {Tolstikhin}, \citenamefont {Morishita},\ and\ \citenamefont
  {Madsen}}]{tolstikhin2011}%
  \BibitemOpen
  \bibfield  {author} {\bibinfo {author} {\bibfnamefont {O.~I.}\ \bibnamefont
  {Tolstikhin}}, \bibinfo {author} {\bibfnamefont {T.}~\bibnamefont
  {Morishita}},\ and\ \bibinfo {author} {\bibfnamefont {L.~B.}\ \bibnamefont
  {Madsen}},\ }\href
  {https://doi.org/https://doi.org/10.1103/PhysRevA.84.053423} {\bibfield
  {journal} {\bibinfo  {journal} {Phys. Rev. A}\ }\textbf {\bibinfo {volume}
  {84}},\ \bibinfo {pages} {053423} (\bibinfo {year} {2011})}\BibitemShut
  {NoStop}%
\bibitem [{\citenamefont {Tong}\ \emph {et~al.}(2002)\citenamefont {Tong},
  \citenamefont {Zhao},\ and\ \citenamefont {Lin}}]{tong2002}%
  \BibitemOpen
  \bibfield  {author} {\bibinfo {author} {\bibfnamefont {X.-M.}\ \bibnamefont
  {Tong}}, \bibinfo {author} {\bibfnamefont {Z.~X.}\ \bibnamefont {Zhao}},\
  and\ \bibinfo {author} {\bibfnamefont {C.-D.}\ \bibnamefont {Lin}},\ }\href
  {https://doi.org/https://doi.org/10.1103/PhysRevA.66.033402} {\bibfield
  {journal} {\bibinfo  {journal} {Phys. Rev. A}\ }\textbf {\bibinfo {volume}
  {66}},\ \bibinfo {pages} {033402} (\bibinfo {year} {2002})}\BibitemShut
  {NoStop}%
\bibitem [{\citenamefont {Zhao}\ \emph {et~al.}(2010)\citenamefont {Zhao},
  \citenamefont {Jin}, \citenamefont {Le}, \citenamefont {Jiang},\ and\
  \citenamefont {Lin}}]{zhao2010}%
  \BibitemOpen
  \bibfield  {author} {\bibinfo {author} {\bibfnamefont {S.-F.}\ \bibnamefont
  {Zhao}}, \bibinfo {author} {\bibfnamefont {C.}~\bibnamefont {Jin}}, \bibinfo
  {author} {\bibfnamefont {A.-T.}\ \bibnamefont {Le}}, \bibinfo {author}
  {\bibfnamefont {T.-F.}\ \bibnamefont {Jiang}},\ and\ \bibinfo {author}
  {\bibfnamefont {C.~D.}\ \bibnamefont {Lin}},\ }\href
  {https://doi.org/https://doi.org/10.1103/PhysRevA.81.033423} {\bibfield
  {journal} {\bibinfo  {journal} {Phys. Rev. A}\ }\textbf {\bibinfo {volume}
  {81}},\ \bibinfo {pages} {033423} (\bibinfo {year} {2010})}\BibitemShut
  {NoStop}%
\bibitem [{\citenamefont {Zhao}\ \emph {et~al.}(2011)\citenamefont {Zhao},
  \citenamefont {Xu}, \citenamefont {Jin}, \citenamefont {Le},\ and\
  \citenamefont {Lin}}]{zhao2011}%
  \BibitemOpen
  \bibfield  {author} {\bibinfo {author} {\bibfnamefont {S.-F.}\ \bibnamefont
  {Zhao}}, \bibinfo {author} {\bibfnamefont {J.}~\bibnamefont {Xu}}, \bibinfo
  {author} {\bibfnamefont {C.}~\bibnamefont {Jin}}, \bibinfo {author}
  {\bibfnamefont {A.-T.}\ \bibnamefont {Le}},\ and\ \bibinfo {author}
  {\bibfnamefont {C.~D.}\ \bibnamefont {Lin}},\ }\href
  {https://doi.org/https://doi.org/10.1088/0953-4075/44/3/035601} {\bibfield
  {journal} {\bibinfo  {journal} {J. Phys. B: At. Mol. Opt. Phys.}\ }\textbf
  {\bibinfo {volume} {44}},\ \bibinfo {pages} {035601} (\bibinfo {year}
  {2011})}\BibitemShut {NoStop}%
\bibitem [{\citenamefont {Kobayashi}\ \emph
  {et~al.}(2020{\natexlab{a}})\citenamefont {Kobayashi}, \citenamefont
  {Neumark},\ and\ \citenamefont {Leone}}]{kobayashi2020}%
  \BibitemOpen
  \bibfield  {author} {\bibinfo {author} {\bibfnamefont {Y.}~\bibnamefont
  {Kobayashi}}, \bibinfo {author} {\bibfnamefont {D.~M.}\ \bibnamefont
  {Neumark}},\ and\ \bibinfo {author} {\bibfnamefont {S.~R.}\ \bibnamefont
  {Leone}},\ }\href
  {https://doi.org/https://doi.org/10.1103/PhysRevA.102.051102} {\bibfield
  {journal} {\bibinfo  {journal} {Phys. Rev. A}\ }\textbf {\bibinfo {volume}
  {102}},\ \bibinfo {pages} {051102(R)} (\bibinfo {year}
  {2020}{\natexlab{a}})}\BibitemShut {NoStop}%
\bibitem [{\citenamefont {Kobayashi}\ \emph
  {et~al.}(2020{\natexlab{b}})\citenamefont {Kobayashi}, \citenamefont {Chang},
  \citenamefont {Poullain}, \citenamefont {Scutelnic}, \citenamefont {Zeng},
  \citenamefont {Neumark},\ and\ \citenamefont {Leone}}]{kobayashi2020a}%
  \BibitemOpen
  \bibfield  {author} {\bibinfo {author} {\bibfnamefont {Y.}~\bibnamefont
  {Kobayashi}}, \bibinfo {author} {\bibfnamefont {K.~F.}\ \bibnamefont
  {Chang}}, \bibinfo {author} {\bibfnamefont {S.~M.}\ \bibnamefont {Poullain}},
  \bibinfo {author} {\bibfnamefont {V.}~\bibnamefont {Scutelnic}}, \bibinfo
  {author} {\bibfnamefont {T.}~\bibnamefont {Zeng}}, \bibinfo {author}
  {\bibfnamefont {D.~M.}\ \bibnamefont {Neumark}},\ and\ \bibinfo {author}
  {\bibfnamefont {S.~R.}\ \bibnamefont {Leone}},\ }\href
  {https://doi.org/https://doi.org/10.1103/PhysRevA.101.063414} {\bibfield
  {journal} {\bibinfo  {journal} {Phys. Rev. A}\ }\textbf {\bibinfo {volume}
  {101}},\ \bibinfo {pages} {063414} (\bibinfo {year}
  {2020}{\natexlab{b}})}\BibitemShut {NoStop}%
\bibitem [{\citenamefont {Xue}\ \emph {et~al.}(2018)\citenamefont {Xue},
  \citenamefont {Du}, \citenamefont {Hu}, \citenamefont {Lin},\ and\
  \citenamefont {Le}}]{xue2018}%
  \BibitemOpen
  \bibfield  {author} {\bibinfo {author} {\bibfnamefont {S.}~\bibnamefont
  {Xue}}, \bibinfo {author} {\bibfnamefont {H.}~\bibnamefont {Du}}, \bibinfo
  {author} {\bibfnamefont {B.}~\bibnamefont {Hu}}, \bibinfo {author}
  {\bibfnamefont {C.~D.}\ \bibnamefont {Lin}},\ and\ \bibinfo {author}
  {\bibfnamefont {A.-T.}\ \bibnamefont {Le}},\ }\href
  {https://doi.org/https://doi.org/10.1103/PhysRevA.97.043409} {\bibfield
  {journal} {\bibinfo  {journal} {Phys. Rev. A}\ }\textbf {\bibinfo {volume}
  {97}},\ \bibinfo {pages} {043409} (\bibinfo {year} {2018})}\BibitemShut
  {NoStop}%
\bibitem [{\citenamefont {Tong}\ \emph {et~al.}(2005)\citenamefont {Tong},
  \citenamefont {Zhao}, \citenamefont {Alnaser}, \citenamefont {Voss},
  \citenamefont {Cocke},\ and\ \citenamefont {Lin}}]{tong2005}%
  \BibitemOpen
  \bibfield  {author} {\bibinfo {author} {\bibfnamefont {X.~M.}\ \bibnamefont
  {Tong}}, \bibinfo {author} {\bibfnamefont {Z.~X.}\ \bibnamefont {Zhao}},
  \bibinfo {author} {\bibfnamefont {A.~S.}\ \bibnamefont {Alnaser}}, \bibinfo
  {author} {\bibfnamefont {S.}~\bibnamefont {Voss}}, \bibinfo {author}
  {\bibfnamefont {C.~L.}\ \bibnamefont {Cocke}},\ and\ \bibinfo {author}
  {\bibfnamefont {C.~D.}\ \bibnamefont {Lin}},\ }\href
  {https://doi.org/https://doi.org/10.1088/0953-4075/38/4/002} {\bibfield
  {journal} {\bibinfo  {journal} {J. Phys. B: At. Mol. Opt. Phys.}\ }\textbf
  {\bibinfo {volume} {38}},\ \bibinfo {pages} {333} (\bibinfo {year}
  {2005})}\BibitemShut {NoStop}%
\bibitem [{\citenamefont {Lam}\ \emph {et~al.}(2020)\citenamefont {Lam},
  \citenamefont {Yarlagadda}, \citenamefont {Venkatachalam}, \citenamefont
  {Wangjam}, \citenamefont {Kushawaha}, \citenamefont {Cheng}, \citenamefont
  {Svihra}, \citenamefont {Nomerotski}, \citenamefont {Weinacht}, \citenamefont
  {Rolles} \emph {et~al.}}]{lam2020}%
  \BibitemOpen
  \bibfield  {author} {\bibinfo {author} {\bibfnamefont {H.~V.~S.}\
  \bibnamefont {Lam}}, \bibinfo {author} {\bibfnamefont {S.}~\bibnamefont
  {Yarlagadda}}, \bibinfo {author} {\bibfnamefont {A.}~\bibnamefont
  {Venkatachalam}}, \bibinfo {author} {\bibfnamefont {T.~N.}\ \bibnamefont
  {Wangjam}}, \bibinfo {author} {\bibfnamefont {R.~K.}\ \bibnamefont
  {Kushawaha}}, \bibinfo {author} {\bibfnamefont {C.}~\bibnamefont {Cheng}},
  \bibinfo {author} {\bibfnamefont {P.}~\bibnamefont {Svihra}}, \bibinfo
  {author} {\bibfnamefont {A.}~\bibnamefont {Nomerotski}}, \bibinfo {author}
  {\bibfnamefont {T.}~\bibnamefont {Weinacht}}, \bibinfo {author}
  {\bibfnamefont {D.}~\bibnamefont {Rolles}}, \emph {et~al.},\ }\href
  {https://doi.org/https://doi.org/10.1103/PhysRevA.102.043119} {\bibfield
  {journal} {\bibinfo  {journal} {Phys. Rev. A}\ }\textbf {\bibinfo {volume}
  {102}},\ \bibinfo {pages} {043119} (\bibinfo {year} {2020})}\BibitemShut
  {NoStop}%
\end{thebibliography}
\end{document}